\documentclass[pdflatex,sn-nature]{sn-jnl}

\usepackage{graphicx}%
\usepackage{multirow}%
\usepackage{amsmath,amssymb,amsfonts}%
\usepackage{amsthm}%
\usepackage[title]{appendix}%
\usepackage{xcolor}%
\usepackage{textcomp}%
\usepackage{manyfoot}%
\usepackage{threeparttable}
\usepackage{booktabs}%
\usepackage[figuresright]{rotating}%
\usepackage{algorithm}%
\usepackage{algorithmicx}%
\usepackage{algpseudocode}%
\usepackage{listings}%
\theoremstyle{thmstyleone}%
\theoremstyle{thmstyletwo}%

\theoremstyle{thmstylethree}%

\begin{document}

\title[Article Title]{Clinician Assessment of Laryngeal Disease with Auditory AI before Laryngoscopy: A Paired Reader Study}


\author[1]{\fnm{Siyuan} \sur{Hou}}\email{housy24@mails.tsinghua.edu.cn}
\equalcont{These authors contributed equally to this work.}

\author[2]{\fnm{Shuting} \sur{Yang}}\email{yst1662@126.com}
\equalcont{These authors contributed equally to this work.}

\author[1]{\fnm{Zengrui} \sur{Jin}}\email{zrjin@tsinghua.edu.cn}
\equalcont{These authors contributed equally to this work.}

\author[2]{\fnm{Liyu} \sur{Cheng}}\email{clystudy@163.com}

\author[1]{\fnm{Siyin} \sur{Wang}}\email{wangsiyi23@mails.tsinghua.edu.cn}


\author[2]{\fnm{Yunhan} \sur{Shi}}\email{yunhanshi@163.com}

\author[2]{\fnm{Xueyan} \sur{Li}}\email{xyli0919@163.com}

\author[2]{\fnm{Rong} \sur{Hu}}\email{hurong\_ronghu@126.com}

\author[2]{\fnm{Yuhong} \sur{Lin}}\email{lyhooooorange@163.com}

\author[2]{\fnm{Qiyu} \sur{Chen}}\email{cqy20010128@163.com}

\author[2]{\fnm{Yanru} \sur{Li}}\email{liyanruru@aliyun.com}

\author[1]{\fnm{Ji} \sur{Wu}}\email{wuji\_ee@tsinghua.edu.cn}

\author[1,3]{\fnm{Bowen} \sur{Zhou}}\email{zbw@pjlab.org.cn}

\author[4]{\fnm{Mangsuo} \sur{Zhao}}\email{zhaomangsuo@163.com}

\author*[2]{\fnm{Wen} \sur{Xu}}\email{xuwendoc@126.com}

\author*[1,3,5]{\fnm{Chao} \sur{Zhang}}\email{cz277@tsinghua.edu.cn}

\affil[1]{\orgdiv{Department of Electronic Engineering}, \orgname{Tsinghua University}, \orgaddress{\city{Beijing}, \postcode{100084}, \country{China}}}

\affil[2]{\orgdiv{Department of Otorhinolaryngology Head and Neck Surgery}, \orgname{Beijing Tongren Hospital, Capital Medical University}, \orgaddress{\city{Beijing}, \postcode{100730}, \country{China}}}

\affil[3]{\orgname{Shanghai Artificial Intelligence Laboratory}, \orgaddress{\city{Shanghai}, \postcode{200030}, \country{China}}}

\affil[4]{\orgdiv{Yuquan Hospital, School of Clinical Medicine, Tsinghua Medicine}, \orgname{Tsinghua University}, \orgaddress{\city{Beijing}, \postcode{100040}, \country{China}}}

\affil[5]{\orgdiv{Department of Human Communication Sciences}, \orgname{University College London}, \orgaddress{\city{London}, \postcode{WC1E 6BT}, \country{United Kingdom}}}


\abstract{

\textbf{Background:} Before laryngoscopy, clinicians form a provisional impression of the likely cause of hoarseness and judge whether a precancerous or malignant lesion should be suspected. Whether voice-based artificial intelligence (AI) supports pre-laryngoscopic categorization is unknown.

\textbf{Methods:} We developed an auditory large language model that generates, from voice recordings, patient information, and acoustic measures obtained before laryngoscopy at one voice center, a patient-level assessment card (provisional category, voice-quality ratings, and rationale) and a prespecified malignancy alert, and evaluated both in a held-out test set and a later cohort. 
In a sequential reader study, 32 otolaryngology clinicians categorized held-out cases without AI and, 1 to 2 weeks later, reviewed their initial categories with the card (without the alert). 
The primary outcome was reader accuracy.

\textbf{Results:} Across 800 paired assessments, reader accuracy was 49.8\% before and 56.8\% after review (difference, 7.0 percentage points; 95\% confidence interval [CI], 1.7 to 13.5; $P<0.001$); 99 categories changed from incorrect to correct and 43 from correct to incorrect (ratio, 2.30; 95\% CI, 1.22 to 5.31).
 Accuracy rose 14.7 percentage points when the model was correct and fell 8.7 when it was incorrect.
 Accuracy of the provisional category was 72\% in the held-out test set and 67\% in temporal validation. The malignancy alert flagged 85.7\% and 86.8\% of precancerous or malignant lesions, whereas the provisional category alone identified 74.3\% and 63.2\%.
 Sensitivity for precancerous or malignant lesions increased from 53.9\% to 75.0\% after AI-assisted review (difference, 21.1 percentage points; 95\% CI, 4.3 to 42.0).

\textbf{Conclusions:} In this single-center study, clinicians' pre-laryngoscopic categorization accuracy was higher after reviewing the AI assessment card, although the sequential design cannot fully separate AI assistance from reassessment; gains depended on model correctness without clearly selective adoption. Randomized and external evaluations, including of the malignancy alert, are needed.

}




\maketitle

\section{Introduction}
\label{sec:intro}

Persistent dysphonia can arise from structural, functional, neurologic, or neoplastic laryngeal disorders whose voice manifestations overlap. 
Establishing the cause requires laryngoscopy, which guidelines recommend when dysphonia persists beyond 4 weeks or whenever a serious cause is suspected \cite{stachler2018clinical}. 
Before the larynx is seen, clinicians form a provisional impression of the likely cause, which may inform ancillary examinations such as stroboscopy for a suspected mucosal lesion, and judge whether a precancerous or malignant lesion should be suspected, which may prompt earlier laryngoscopy or narrow-band imaging. The second judgment tolerates fewer misses than the first.
Both judgments are unreliable even among specialists: six laryngologists given only the history and physical examination reached the correct diagnosis in 5\% of cases and identified cancer in 33\% \cite{https://doi.org/10.1002/lary.23630}. 
Voice samples were not provided in that study, and how much the voice adds is not well quantified.
In otolaryngology and voice clinics, the voice is routinely documented by a standardized assessment (recordings of sustained vowels and read speech, with acoustic measures such as jitter and shimmer) that does not depend on laryngoscopic findings \cite{dejonckere2001basic, patel2018recommended}, which offers a defined point at which decision support could be delivered.

Voice-based artificial intelligence (AI) has been studied for laryngeal disease classification and automated perceptual voice rating \cite{fagherazzi2021voice,kim2024classification,berteloot2026ai,hidaka2025automatic}, but published evaluations have largely been limited to standalone algorithm performance \cite{mairesse2026diagnostic}.
Standalone accuracy does not establish how a model's output will change clinicians' decisions.
In other diagnostic tasks, AI assistance has improved clinician performance \cite{barnett2024clinician}, failed to improve it despite strong standalone performance \cite{goh2024llm}, and worsened it when the AI recommendation was wrong \cite{qazi2026automation}.
Because the net effect depends on whether the errors of clinicians and models are complementary \cite{zwaan2026trust}, average assisted accuracy can conceal how often assistance corrects an initially incorrect decision and how often it overturns a correct one.
Whether voice-based AI helps or harms clinicians' pre-laryngoscopic judgments is unknown.


\begin{figure}
	\centering
	\includegraphics[width=\linewidth]{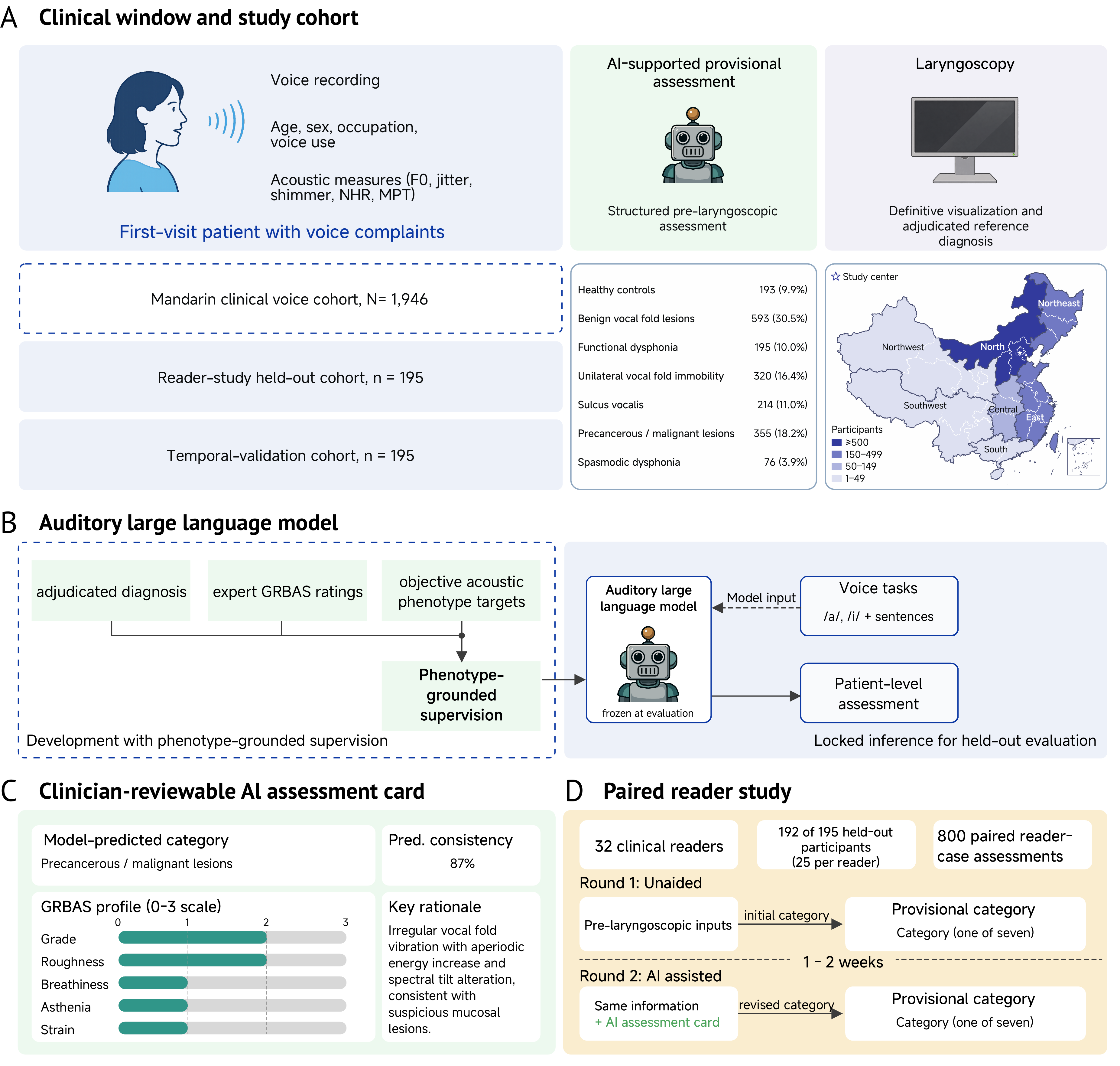}
	\caption{\textbf{Study Design.} Panel A shows the intended pre-laryngoscopic point of use, after a standardized voice assessment, and distinguishes the single-center development cohort (partitioned into training, validation, and held-out test sets) from the subsequent temporal-validation cohort. Panel B shows model development and evaluation: during training, target responses (category, GRBAS scores, and rationale) were constructed from the reference category, reference GRBAS ratings, and acoustic measures; at evaluation, the locked model received only the pre-laryngoscopic inputs and produced the assessment card and a separate risk-flagging malignancy alert. Panel C shows an example of the assessment card, translated into English. Panel D shows the two-round sequential reader study conducted on held-out test cases using the assessment card (without the malignancy alert).
}
\label{fig:1}
\end{figure}


We developed an auditory large language model \cite{wang2026general} (a large language model that accepts audio as direct input) that generates two outputs from the routine pre-laryngoscopic voice assessment: an assessment card comprising a provisional category (one of seven), voice-quality ratings on the GRBAS scale (grade, roughness, breathiness, asthenia, and strain, each 0 to 3) \cite{hirano1981psychoacoustic}, a brief rationale, and the agreement among the model's readings of the patient's recordings; and a malignancy alert to flag suspected precancerous or malignant lesions, prespecified to favor sensitivity (Fig. \ref{fig:1}). 
We evaluated both outputs, with the model fixed, in a held-out sample from a single-center cohort of Mandarin-speaking participants and in participants assessed later at the same center. 
In a sequential reader study, 32 otolaryngology clinicians, from trainees to chief physicians, categorized the held-out cases without AI and, 1 to 2 weeks later, reviewed their initial categories with the card, which carried the provisional category but not the alert.
We examined whether categorization accuracy was higher after review and how often initial categories changed from incorrect to correct or the reverse; downstream examination and referral decisions were not measured.

\section{Methods}\label{sec:methods}

\subsection{Study Design and Oversight}
\label{sec:methods-design}

%
%

This study comprised development and held-out evaluation of an auditory large language model in a retrospective single-center cohort, temporal validation of the locked model in participants assessed later at the same center, and a reader study in which clinicians categorized held-out cases without and then with the model's assessment card. 
Model weights, inference procedures, and the malignancy alert were fixed with the use of training and validation data only, before held-out testing and reader-study data collection. 
The institutional review board of Beijing Tongren Hospital approved the retrospective use of recordings and clinical data with a waiver of informed consent and the reader study (approval no. TREC2026-KY282).
Readers provided informed consent.

\subsection{Participants, Voice Assessment, and Reference Standard}
\label{sec:methods-participants}

Participants were drawn from routine assessments at the Voice Center of Beijing Tongren Hospital between October 2004 and March 2026 (Table \ref{tab:laryngeal_cohort_year_distribution}).
Patients were eligible if hoarseness was the presenting complaint and the final diagnosis was benign vocal fold lesions, functional dysphonia, unilateral vocal fold immobility, sulcus vocalis, precancerous or malignant lesions, or spasmodic dysphonia.
Patients with coexisting voice disorders, previous vocal fold surgery, or insufficient documentation were excluded. 
Healthy controls were patients without voice complaints who had normal findings on laryngoscopy before benign thyroid surgery, recorded with the same equipment. 
Participants were partitioned 8:1:1 into training, validation, and held-out test sets, stratified according to category and sex, with each participant's recordings in one set. 
The temporal-validation cohort comprised all eligible participants assessed from April through August 2026; none contributed to model development.

The voice recordings (sustained /a/ and /i/ and a standardized Mandarin sentence), patient information, and routine acoustic measures (Fig. \ref{fig:1}A and Supplementary Section \ref{sec:supp_methods_recording}) were all obtained before the larynx was visualized; we refer to them as the pre-laryngoscopic inputs. 
The model and the readers received only these inputs; missing fields were omitted rather than imputed.

The reference category was assigned by a board-certified otolaryngologist on the basis of clinical assessment and laryngostroboscopy, with narrow-band imaging when indicated and histopathological confirmation for precancerous or malignant lesions.
Inconsistent cases were adjudicated by three senior otolaryngologists unaware of the model output, and acoustic analysis was not used. 
Reference GRBAS ratings \cite{hirano1981psychoacoustic} were assigned at the voice assessment by a trained voice clinician and reviewed by a senior voice clinician.

\subsection{AI System} 
\label{sec:methods-model} 

%


The model was built on an open-weight auditory large language model (Qwen3-Omni-30B-A3B-Thinking, a sparse mixture-of-experts model activating 3 billion of 30 billion parameters per token) \cite{qwen2025omni} adapted by low-rank adaptation \cite{lora} and fine-tuned jointly on seven-class categorization, GRBAS rating, and related speech-assessment tasks. 
A single checkpoint was selected according to patient-level weighted F1 in the validation set (Supplementary Section \ref{sec:supp_methods_training}). 
For each participant, the model was applied to every recording and to every combination of that participant's recordings joined end to end (seven readings for the three standard recordings; median, 7 [range, 1 to 63] in the held-out test set); each application, which we call a reading, yielded a category, GRBAS scores, and a rationale (Supplementary Section \ref{sec:supp_methods_inference}). 
From these readings the model produced two patient-level outputs. 
The provisional category was the category favored by most readings, GRBAS scores were the means across readings, and agreement among readings was the proportion of readings favoring the provisional category (reflecting within-patient stability across overlapping recordings rather than a calibrated probability). 
%
The malignancy alert was raised when the proportion of valid readings favoring precancerous or malignant lesions reached at least 2/7 (approximately 28.6\%), the highest threshold that maintained a sensitivity of at least 80\% in the validation set (Supplementary Section \ref{sec:supp_methods_alert}).
The assessment card displayed the provisional category, the number of readings favoring each category, the GRBAS scores, agreement among readings, and one rationale from a reading that matched the provisional category (Fig. \ref{fig:1}C); generated rationales were not independently audited for clinical faithfulness at inference, and the alert was not displayed.

\subsection{Standalone Evaluation and Temporal Validation}
\label{sec:methods-standalone}

Standalone performance was assessed in the held-out test set against the reference category, with patient-level weighted F1 for the provisional category as the principal measure and accuracy reported alongside. 
Classical acoustic classifiers \cite{svm,kim2024classification,gmm_i_vector,lad2019total,gmm_hmm,mesallam2017development}, fine-tuned speech encoders \cite{wav2vec2,cai2024voice,tirronen2023hierarchical,whisper,merler2025clinical}, and a single-task version of the model trained on the same partitions served as comparators (Supplementary Section \ref{sec:supp_methods_comparators}); because available non-audio inputs differed across these pipelines, comparisons reflect implemented systems rather than strictly input-matched architectures. Model-generated GRBAS scores were compared with the reference ratings. 
Recognition of precancerous or malignant lesions was evaluated with the malignancy alert and, for comparison, with the provisional category alone. 
The area under the receiver-operating-characteristic curve (AUROC) was calculated from the proportion of readings favoring these lesions. 
In the temporal-validation cohort, the model and the alert were applied without modification.

\subsection{Reader Study}
\label{sec:reader_study}

Thirty-two otolaryngology clinicians, from postgraduate trainees to chief physicians, were recruited from hospitals in Beijing and elsewhere in China (Table \ref{tab:reader_characteristics}). 
Each reader was independently assigned 25 participants sampled at random without replacement from the 195-participant held-out test set, and participants could be assigned to more than one reader.
In the first round, readers listened to the recordings, reviewed the other pre-laryngoscopic inputs, and selected one of the seven categories. 
In the second round, 1 to 2 weeks later, readers evaluated the same participants and were shown the same information, their own first-round category, and the assessment card, and retained or changed the category.
The fixed order of rounds reflects the intended use of the card as a check on the clinician's initial category and preserves an unaided baseline for every reader and case.
Case order was randomized in each round, and readers were unaware of the reference category and of other readers' responses.

A response was counted as correct if it matched the reference category. 
The primary outcome was accuracy, the proportion of correct reader--case assessments, in each round. 
Secondary outcomes were macro-averaged recall and F1, transitions between rounds (incorrect to correct and correct to incorrect), and reader sensitivity and specificity for precancerous or malignant lesions, with a response counted as positive when that category was selected. Analyses stratified according to model correctness and reader or case characteristics were exploratory.

\subsection{Statistical Analysis}
\label{sec:methods-statistical}

Confidence intervals for standalone measures were percentile intervals from 5000 bootstrap resamples of participants. 
Agreement between model-generated and reference GRBAS scores was summarized with the mean absolute error and the intraclass correlation coefficient \cite{koo2016guideline}. 

The primary analysis was a mixed-effects logistic regression of correct categorization on study round, with random intercepts for reader and participant. 
The between-round difference is summarized as an odds ratio with a 95\% confidence interval and a two-sided P value. 
Confidence intervals for between-round differences and for the ratio of incorrect-to-correct to correct-to-incorrect transitions were obtained by bootstrap resampling of both readers and participants, with paired responses kept together. 
Sensitivity analyses addressed pair-level correlation across rounds and reader-specific round slopes using alternative random-effects structures, including a reader--participant dyad intercept (Table \ref{tab:reader_primary_robustness}). 
Exploratory heterogeneity was assessed by adding each factor and its interaction with study round to the primary model.
The primary analysis was the sole prespecified confirmatory hypothesis test; all interaction analyses were exploratory. Confidence intervals for secondary and exploratory outcomes were not adjusted for multiplicity and should not be used to infer definitive effects.
No formal sample-size calculation was performed; the sample was determined by the available records and participating readers.
Analyses were performed with Python 3.10 and R 4.5.3 (Supplementary Section \ref{sec:supp_methods_statistics}).

\section{Results}\label{sec:results}

\subsection{Participants and Study Cohorts}
\label{sec:results-cohort}

\begin{table}[htbp]
\centering
\setlength{\tabcolsep}{4pt}
\caption{Characteristics of the Participants.}
\label{tab:cohort_characteristics}
\begin{tabular}{@{}lcccc@{}}
\toprule
 & \multicolumn{3}{c}{Development cohort} & Temporal \\
\cmidrule(lr){2-4}
Characteristic & Training & Validation & Held-out test & validation \\
\midrule
\multicolumn{5}{@{}l}{\textit{Cohort size}} \\
\quad Participants, no. & 1556 & 195 & 195 & 195 \\
\quad Recordings, no. & 4698 & 600 & 586 & 605 \\
\quad Recordings per participant, median (IQR) & 3 (3--3) & 3 (3--3) & 3 (3--3) & 3 (3--3) \\
\midrule
\multicolumn{5}{@{}l}{\textit{Demographics}} \\
\quad Age, yr, median (IQR) & 49 (35--60) & 51 (37--61) & 49 (36--60) & 49 (36--58) \\
\quad Female sex, no. (\%) & 732 (47.0) & 91 (46.7) & 99 (50.8) & 91 (46.7) \\
\midrule
\multicolumn{5}{@{}l}{\textit{Reference category, no. (\%)}} \\
\quad Healthy control & 154 (9.9) & 19 (9.7) & 20 (10.3) & 22 (11.3) \\
\quad Benign vocal fold lesions & 474 (30.5) & 59 (30.3) & 60 (30.8) & 63 (32.3) \\
\quad Functional dysphonia & 156 (10.0) & 20 (10.3) & 19 (9.7) & 21 (10.8) \\
\quad Unilateral vocal fold immobility & 256 (16.5) & 32 (16.4) & 32 (16.4) & 33 (16.9) \\
\quad Sulcus vocalis & 171 (11.0) & 21 (10.8) & 22 (11.3) & 12 (6.2) \\
\quad Precancerous or malignant lesions & 284 (18.3) & 36 (18.5) & 35 (17.9) & 38 (19.5) \\
\quad Spasmodic dysphonia & 61 (3.9) & 8 (4.1) & 7 (3.6) & 6 (3.1) \\
\bottomrule
\end{tabular}
\end{table}

The development cohort comprised 1946 participants, of whom 195 formed the held-out test set; the temporal-validation cohort comprised a further 195 participants (Table \ref{tab:cohort_characteristics}). 
The distribution of categories was similar across sets, except that sulcus vocalis was less common in the temporal-validation cohort than in the held-out test set (6.2\% vs. 11.3\%). 
Precancerous or malignant lesions were present in 35 participants (17.9\%) and 38 (19.5\%), respectively.

All 32 readers completed both rounds, yielding 800 paired reader--case assessments with no missing responses. The independently randomized assignments included 192 of the 195 held-out participants; each included participant was assessed by a median of 4 readers (range, 1 to 11).
Readers comprised 6 postgraduate trainees, 10 residents or junior physicians, 10 attending or intermediate physicians, and 6 chief physicians; 14 practiced outside Beijing (Table \ref{tab:reader_characteristics}).

\subsection{Standalone Performance}
\label{sec:results-standalone}

\begin{figure}[htbp]
\centering
\includegraphics[width=\linewidth]{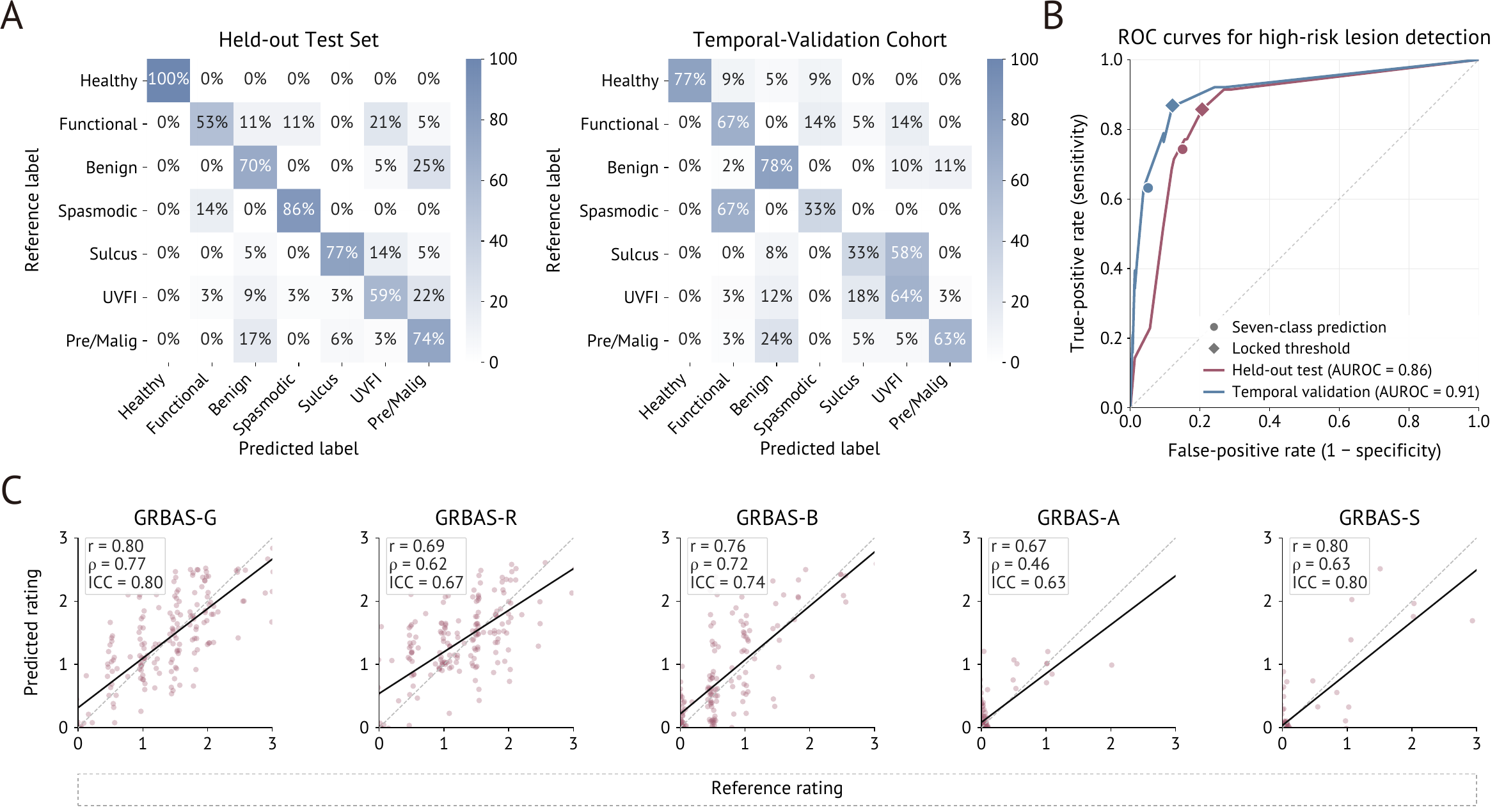}
\caption{
\textbf{Standalone Performance in the Held-out Test Set and the Temporal-Validation Cohort.} 
Panel A shows confusion matrices for seven-class categorization (rows, reference category; columns, predicted category). UVFI denotes unilateral vocal fold immobility, and Pre/Malig denotes precancerous or malignant lesions.
Panel B shows receiver-operating-characteristic curves for precancerous or malignant lesions, based on the proportion of the model's readings favoring that category; diamonds indicate the malignancy alert, and circles the provisional category alone. 
Panel C shows patient-level agreement between model-generated and reference GRBAS scores in the held-out test set.
}
\label{fig:2}
\end{figure}

In the held-out test set, the provisional category was correct for 140 of 195 participants (accuracy, 0.72; 95\% CI, 0.65 to 0.78), and patient-level weighted F1 was 0.72 (95\% CI, 0.66 to 0.78). 
Among the six disease categories, sensitivity ranged from 53\% for functional dysphonia to 86\% for spasmodic dysphonia; all 20 healthy controls were classified correctly (Fig. \ref{fig:2}A).
Weighted F1 was numerically higher than that of a single-task version of the model (0.68) and comparator speech-encoder (0.62) and classical acoustic (0.53) pipelines (Fig. \ref{fig:supp_laryngeal_standalone}B and Table \ref{tab:laryngeal_etiology_baseline_comparison}).
At the patient level, the mean absolute error for GRBAS scores averaged across the five dimensions was 0.25 points, and intraclass correlation coefficients ranged from 0.63 to 0.80 (Fig. \ref{fig:2}C).

The malignancy alert was raised for 30 of the 35 participants with precancerous or malignant lesions, missing 5 (sensitivity, 85.7\%; specificity, 79.4\%; AUROC, 0.86) (Table \ref{tab:temporal_validation} and Fig. \ref{fig:2}B). 
Had the provisional category alone been used to flag these lesions, 26 of 35 would have been identified (sensitivity, 74.3\%; specificity, 85.0\%).

\subsection{Temporal Validation} 
\label{sec:results-temporal} 

In the temporal-validation cohort, accuracy of the provisional category was 0.67 (95\% CI, 0.61 to 0.74), weighted F1 was 0.68 (95\% CI, 0.61 to 0.74), and macro-averaged F1 was 0.60 (Table \ref{tab:temporal_validation}); sensitivity was lowest for sulcus vocalis (4 of 12 participants) and spasmodic dysphonia (2 of 6) (Table \ref{tab:temporal_validation_class_performance}). The malignancy alert was raised for 33 of 38 participants with precancerous or malignant lesions, missing 5 (sensitivity, 86.8\%; specificity, 87.9\%; AUROC, 0.91), whereas the provisional category alone would have identified 24 of 38 (sensitivity, 63.2\%) (Table \ref{tab:temporal_validation}).


\begin{table}[htbp]
\centering
\setlength{\tabcolsep}{4pt}
\caption{Standalone Performance in the Held-out Test Set and the Temporal-Validation Cohort.}
\label{tab:temporal_validation}
\begin{tabular}{@{}lcc@{}}
\toprule
Measure & Held-out test set (N\,=\,195) & Temporal validation (N\,=\,195) \\
\midrule
\multicolumn{3}{@{}l}{\textit{Provisional category (seven classes)}} \\
\quad Accuracy & 0.72 (0.65 to 0.78) & 0.67 (0.61 to 0.74) \\
\quad Weighted F1 & 0.72 (0.66 to 0.78) & 0.68 (0.61 to 0.74) \\
\quad Macro-averaged F1 & 0.74 (0.66 to 0.80) & 0.60 (0.51 to 0.67) \\
\midrule
\multicolumn{3}{@{}l}{\textit{Precancerous or malignant lesions: provisional category alone}} \\
\quad Sensitivity, \% & 74.3 (59.3 to 88.5) & 63.2 (47.2 to 78.6) \\
\quad Specificity, \% & 85.0 (79.3 to 90.3) & 94.9 (91.2 to 98.1) \\
\quad Lesions missed, no./total no. & 9/35 & 14/38 \\
\midrule
\multicolumn{3}{@{}l}{\textit{Precancerous or malignant lesions: malignancy alert}} \\
\quad AUROC & 0.86 (0.79 to 0.92) & 0.91 (0.85 to 0.96) \\
\quad Sensitivity, \% & 85.7 (72.7 to 96.7) & 86.8 (75.0 to 97.1) \\
\quad Specificity, \% & 79.4 (73.2 to 85.3) & 87.9 (82.8 to 92.9) \\
\quad Positive predictive value, \% & 47.6 (35.2 to 60.0) & 63.5 (50.0 to 76.5) \\
\quad Negative predictive value, \% & 96.2 (92.6 to 99.2) & 96.5 (93.1 to 99.3) \\
\quad Lesions missed, no./total no. & 5/35 & 5/38 \\
\bottomrule
\end{tabular}
\footnotetext{Values in parentheses are 95\% confidence intervals from 5000 bootstrap resamples of participants.
The provisional category alone was counted as positive when it was precancerous or malignant lesions.
The malignancy alert was raised when the proportion of the model's readings favoring that category was at least 2/7; this proportion was set in the validation set and fixed before held-out testing.
AUROC denotes area under the receiver-operating-characteristic curve, calculated from the same proportion.
Precancerous or malignant lesions were present in 35 participants (17.9\%) in the held-out test set and in 38 (19.5\%) in the temporal-validation cohort; predictive values depend on this prevalence.}
\end{table}

\subsection{Reader Performance Before and After Review with the Card}
\begin{figure}[htbp]
\centering
\includegraphics[width=0.98\linewidth]{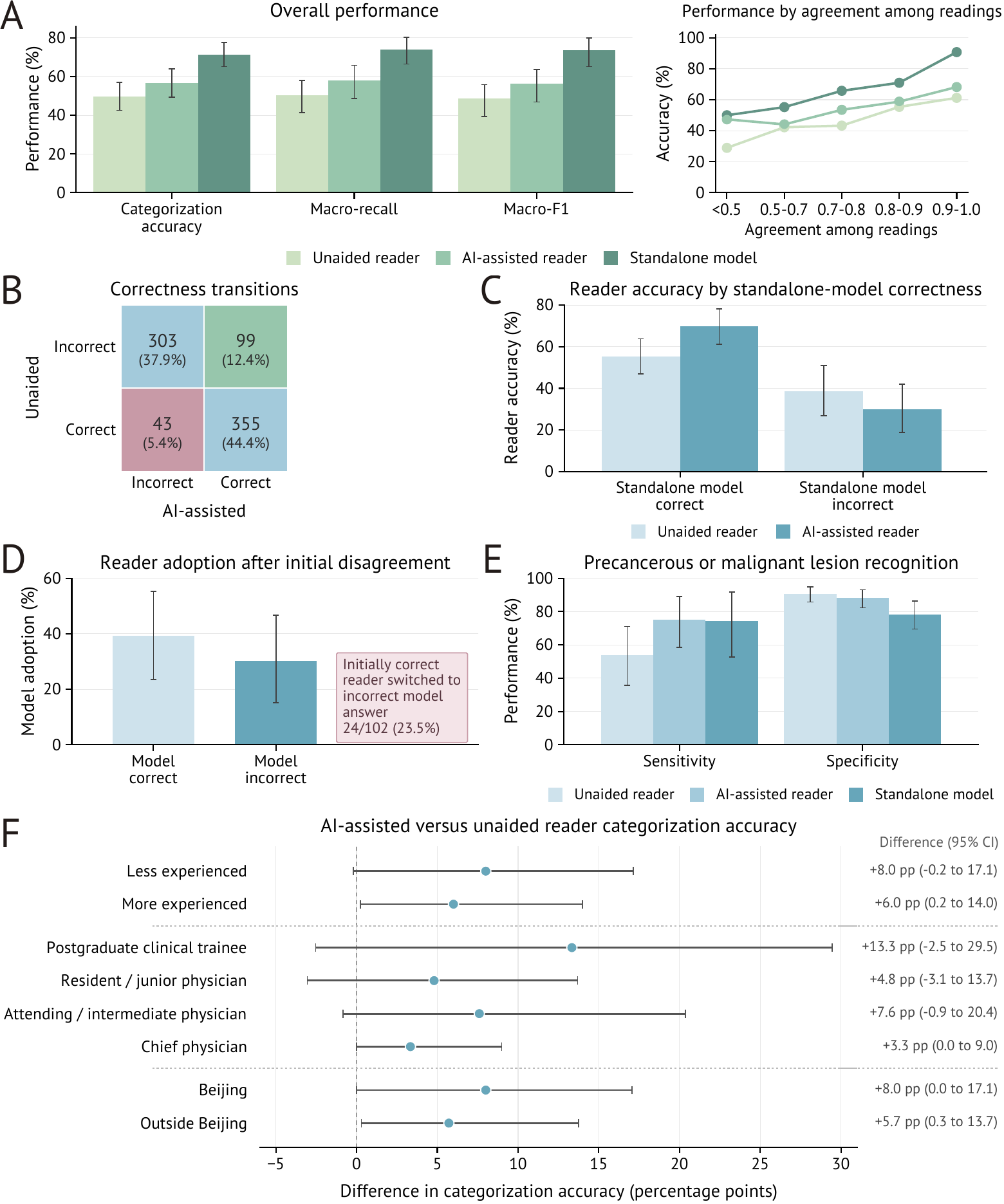}
\caption{
\textbf{Reader Performance Before and After Review with the Assessment Card.}
Panel A shows accuracy, macro-averaged recall, and macro-averaged F1 across 800 paired assessments and accuracy according to agreement among readings; standalone-model estimates use the 192 unique participants represented in the reader study.
Panel B shows paired transitions between rounds. 
Panel C shows reader accuracy according to whether the model was correct. 
Panel D shows the proportion of assessments in which readers who initially disagreed with the model adopted its category. 
Panel E shows reader sensitivity and specificity for precancerous or malignant lesions (34 participants, 128 assessments), based on the category selected; standalone-model estimates in this panel use the same reader--case weighting.
Panel F shows the between-round difference in accuracy according to reader characteristics.
Error bars indicate 95\% confidence intervals, which were not adjusted for multiplicity.
}
\label{fig:3}
\end{figure}

Reader accuracy was 49.8\% in the initial unaided round and 56.8\% after review with the card (difference, 7.0 percentage points; 95\% CI, 1.7 to 13.5; odds ratio, 1.49; 95\% CI, 1.18 to 1.87; $P<0.001$) (Fig. \ref{fig:3}A).
The association remained in a sensitivity analysis that included a random intercept for each reader--participant dyad (odds ratio, 1.92; 95\% CI, 1.49 to 2.49; Table \ref{tab:reader_primary_robustness}).
Macro-averaged recall and F1 increased by 7.5 (95\% CI, 1.0 to 15.1) and 7.6 (95\% CI, 1.6 to 14.7) percentage points, respectively.
Among the 192 unique participants represented in the reader study, standalone model accuracy was 71.4\%.

Of the 800 paired assessments, 355 (44.4\%) were correct in both rounds and 303 (37.9\%) were incorrect in both; 99 (12.4\%) changed from incorrect to correct and 43 (5.4\%) from correct to incorrect, a ratio of 2.30 (95\% CI, 1.22 to 5.31) (Fig. \ref{fig:3}B).

For precancerous or malignant lesions (128 assessments of 34 participants), with a card that carried the provisional category but not the alert, reader sensitivity was 53.9\% before and 75.0\% after review (difference, 21.1 percentage points; 95\% CI, 4.3 to 42.0), and specificity was 90.6\% and 88.1\% (difference, $-$2.5 percentage points; 95\% CI, $-$7.3 to 2.0) (Fig. \ref{fig:3}E).





\subsection{Reader Performance According to Model Correctness and Reader Characteristics}
In exploratory analyses stratified by model correctness, the model was correct for 137 participants (536 assessments) and incorrect for 55 (264 assessments).
When the model was correct, reader accuracy was 55.2\% before and 70.0\% after review (difference, 14.7 percentage points; 95\% CI, 6.8 to 23.6); when the model was incorrect, accuracy was 38.6\% and 29.9\%, respectively (difference, $-$8.7 percentage points; 95\% CI, $-$18.6 to $-$1.2). The interaction odds ratio was 3.86 (95\% CI, 2.33 to 6.39; Holm-adjusted $P<0.001$); corresponding results are shown in Fig.~\ref{fig:3}C and Tables~\ref{tab:reader_model_correctness} and~\ref{tab:reader_model_correctness_transitions}.

In 429 assessments, the reader's unaided category differed from the model's.
Readers then adopted the model's category in 94 of 240 assessments (39.2\%) in which the model was correct and in 57 of 189 (30.2\%) in which it was incorrect (odds ratio, 1.82; 95\% CI, 1.01 to 3.27; Holm-adjusted $P=0.147$) (Fig. \ref{fig:3}D).
In 102 assessments, the reader was initially correct and the model was not; in 24 of these (23.5\%), the reader switched to the model's category.

The difference in accuracy between rounds was 8.0 percentage points among less-experienced readers and 6.0 among more-experienced readers (interaction odds ratio, 0.90; 95\% CI, 0.57 to 1.42), and it ranged from 13.3 percentage points among postgraduate trainees to 3.3 among chief physicians (Fig. \ref{fig:3}F).
The difference did not vary appreciably with agreement among readings (interaction odds ratio per 0.1 increase, 0.97; 95\% CI, 0.86 to 1.09), practice region, or reference category (Tables \ref{tab:reader_exploratory_interactions} and \ref{tab:reader_disease_specific}).

\section{Discussion}
\label{sec:discussion}

In this single-center study, an auditory large language model generated a patient-level assessment card from a standardized voice assessment, and among 32 otolaryngology clinicians across China who reviewed their initial pre-laryngoscopic categorization with the card, accuracy was higher after review than before (56.8\% vs.\ 49.8\%), without clear heterogeneity across reader experience or practice region.
Accuracy rose when the model was correct and fell when it was incorrect, and model adoption was not clearly higher for correct than for incorrect recommendations.
Evaluated without clinicians, a prespecified malignancy alert flagged approximately 86\% of precancerous or malignant lesions in both the held-out test set and a later cohort from the same center, whereas the accuracy of the provisional category was lower in the later cohort, particularly for the least common categories.

These findings characterize how clinicians use voice-based AI output during sequential review. 
The net gain arose because the model was correct more often than not, not because readers reliably identified when to rely on it. Among initial disagreements, readers adopted the model's category in 39\% of assessments when it was correct and 30\% when it was incorrect, providing no robust evidence of selective reliance. This pattern also helps explain why accuracy after review remained below that of the model alone.
Although the fixed-order design cannot fully separate the influence of the card from second-look reassessment or anchoring, two observations suggest that reassessment alone does not explain these shifts. 
Readers changed only 5.1\% of initially correct categories with which the model agreed, an estimate of background revision when reader and card concurred, whereas 23.5\% of initially correct readers switched to an incorrect model category, and accuracy fell when the model was wrong, a bidirectional pattern unlikely to arise from practice or recall alone.
Similar patterns have been reported in other diagnostic tasks \cite{barnett2024clinician, goh2024llm, qazi2026automation, gaube2021ai}. 
The aim of decision support is appropriate reliance \cite{zwaan2026trust, yu2024heterogeneity}, and the between-round difference did not vary with displayed agreement among readings, which reflects stability across overlapping recordings rather than a calibrated probability. 
Explanations produced by clinical speech AI, including the card's unverified rationale, have seldom been validated for clinical faithfulness \cite{ebraheem2026explainable}. 
In other settings, explanations have increased acceptance of AI recommendations regardless of their correctness \cite{chanda2024} and have not protected clinicians from the errors of a biased model \cite{jabbour2023}.

The two model outputs serve distinct roles, and the reader study tested only the provisional category on the card, which alone identified 74\% and 63\% of precancerous or malignant lesions in the two cohorts.
Even without the malignancy alert, reader sensitivity for precancerous or malignant lesions rose from 54\% before review to 75\% after review, whereas specificity changed from 90.6\% to 88.1\%.
The malignancy alert flagged 86\% of such lesions in both cohorts, but still missed roughly 14\% (10 lesions across the two sets); accordingly, neither output can be used to exclude malignancy or defer laryngoscopy, and the alert is strictly a candidate risk flag rather than a rule-out test. Whether adding the alert to the card alters clinician prioritization or referral decisions remains untested.

This study has limitations. The fixed-order sequential design cannot fully separate the influence of the card from repeat assessment or anchoring to the initial category. 
Development, temporal validation, and reader testing were conducted with data from one specialty center, in a Mandarin-speaking cohort restricted to six disease categories and asymptomatic controls, with otolaryngology clinicians as readers; generalizability to external hospitals, unselected patients with hoarseness, and nonspecialists is unknown. 
Comparator baselines differed in non-audio inputs, and a baseline using patient context and routine acoustic measures without audio was not evaluated, so the incremental contribution of raw audio alone was not isolated.
We measured categorization accuracy, a surrogate for downstream examination and referral decisions, rather than those decisions or patient outcomes; the malignancy alert was evaluated only without clinicians; and multi-reading inference was run offline on server GPUs without deployment optimization, leaving clinical workflow latency and resource requirements to be established.

%
In conclusion, otolaryngology clinicians’ pre-laryngoscopic categorization accuracy was higher after reviewing an assessment card from an auditory large language model. The net benefit depended strongly on model correctness, and the sequential design could not exclude reassessment effects. A randomized, multicenter study with an unaided re-review control is needed to evaluate the assessment card and malignancy alert.

\section*{Declarations}\label{sec:declarations}

\subsection*{Disclosures}

The authors declare no competing interests.

\subsection*{Data and Code Availability}

The clinical voice cohort contains protected health information and is therefore not publicly available.
De-identified summary data, reader-level metric tables and adjudicated reference labels for the reader-study set will be made available to qualified researchers under a data-use agreement on reasonable request to the corresponding author and subject to approval by the institutional review board of the originating site(s).

The training and evaluation code for the auditory large language model and the reader-study analysis scripts will be made publicly available upon publication, with model weights released under a research-use license.

\subsection*{Author Contributions}

Siyuan Hou and Siyin Wang developed the model.
Siyuan Hou performed the statistical analyses and prepared the figures.
Shuting Yang contributed to study design, collected the clinical data, performed reference annotation and quality control, and organized the reader study.
Liyu Cheng, Yunhan Shi, Xueyan Li, Rong Hu, Yuhong Lin, and Qiyu Chen collected the clinical data.
Zengrui Jin wrote the manuscript.
Yanru Li, Ji Wu, Bowen Zhou, and Mangsuo Zhao contributed to manuscript preparation.
Wen Xu and Chao Zhang conceived and supervised the study, leading the clinical and AI components, respectively. 

\subsection*{Acknowledgments}

We thank the physician readers, voice clinicians and patients who made this study possible.

OpenAI's ChatGPT was used to assist with language editing and editorial refinement of the manuscript. The authors reviewed all AI-assisted content and take full responsibility for its accuracy and integrity.

\bibliography{sn-bibliography,bib/intro,bib/method}

@article{fagherazzi2021voice,
  title={Voice for health: the use of vocal biomarkers from research to clinical practice},
  author={Fagherazzi, Guy and Fischer, Aur{\'e}lie and Ismael, Muhannad and Despotovic, Vladimir},
  journal={Digital biomarkers},
  volume={5},
  number={1},
  pages={78--88},
  year={2021},
  publisher={S. Karger AG Basel, Switzerland}
}

@article{stachler2018clinical,
  title={Clinical Practice Guideline: Hoarseness ({D}ysphonia) ({U}pdate)},
  author={Stachler, Robert J and Francis, David O and Schwartz, Seth R and Damask, Cecelia C and Digoy, German P and Krouse, Helene J and McCoy, Scott J and Ouellette, Daniel R and Patel, Rita R and Reavis, Charles Charlie W and Smith, Libby J and Smith, Marshall and Strode, Steven W and Woo, Peak and Nnacheta, Lorraine C},
  journal={Otolaryngology--Head and Neck Surgery},
  volume={158},
  number={1\_suppl},
  pages={S1--S42},
  year={2018},
  doi={10.1177/0194599817751030}
}

@article{kim2024classification,
  title={Classification of laryngeal diseases including laryngeal cancer, benign mucosal disease, and vocal cord paralysis by artificial intelligence using voice analysis},
  author={Kim, Hyun-Bum and Song, Jaemin and Park, Seho and Lee, Yong Oh},
  journal={Scientific reports},
  volume={14},
  number={1},
  pages={9297},
  year={2024},
  publisher={Nature Publishing Group UK London}
}

@article{wang2026general,
  title={Towards general auditory intelligence for machine listening and speaking},
  author={Wang, Siyin and Jin, Zengrui and Tang, Changli and Li, Qiujia and Li, Bo and Chen, Chen and Hu, Yuchen and Yu, Wenyi and Li, Yixuan and Zhuang, Jimin and Yang, Yudong and Wang, Mingqiu and Han, Michael and Ding, Yifan and Bai, Junwen and Ouyang, Tom and Chang, Shuo-yiin and Chen, Xianzhao and Tian, Xiaohai and Zhang, Jun and Lu, Lu and Sun, Guangzhi and Chen, Zhehuai and Wu, Ji and Zhou, Bowen and Wang, Yuxuan and Sainath, Tara and Wu, Yonghui and Zhang, Chao},
  journal={Nature Machine Intelligence},
  year={2026},
  volume={8},
  pages={1201--1220},
  doi={10.1038/s42256-026-01281-1}
}

@misc{qwen2025omni,
  title = {Qwen3-Omni Technical Report},
  author = {{Qwen Team}},
  year = {2025},
  howpublished = {arXiv preprint arXiv:2509.17765}
}

@article{hidaka2025automatic,
  title   = {Automatic {GRBAS} Scoring of Pathological Voices using Deep Learning and a Small Set of Labeled Voice Data},
  author  = {Hidaka, Shunsuke and Lee, Yogaku and Nakanishi, Moe and Wakamiya, Kohei and Nakagawa, Takashi and Kaburagi, Tokihiko},
  journal = {Journal of Voice},
  volume  = {39},
  number  = {3},
  pages   = {846.e1--846.e23},
  year    = {2025},
  doi     = {10.1016/j.jvoice.2022.10.020}
}

@article{ebraheem2026explainable,
  title={Explainable and Interpretable {AI} for Voice and Speech Analysis in Clinical Care: Systematic Review},
  author={Ebraheem, Mohamed and Toghranegar, Jamie and {Bridge2AI-Voice Consortium} and Bensoussan, Yael and Templeton, John Michael},
  journal={Journal of Medical Internet Research},
  volume={28},
  pages={e83790},
  year={2026},
  doi={10.2196/83790}
}

@article{yu2024heterogeneity,
  title={Heterogeneity and predictors of the effects of AI assistance on radiologists},
  author={Yu, Feiyang and Moehring, Alex and Banerjee, Oishi and Salz, Tobias and Agarwal, Nikhil and Rajpurkar, Pranav},
  journal={Nature Medicine},
  volume={30},
  number={3},
  pages={837--849},
  year={2024},
  publisher={Nature Publishing Group US New York}
}

@article{gaube2021ai,
  title={Do as AI say: susceptibility in deployment of clinical decision-aids},
  author={Gaube, Susanne and Suresh, Harini and Raue, Martina and Merritt, Alexander and Berkowitz, Seth J and Lermer, Eva and Coughlin, Joseph F and Guttag, John V and Colak, Errol and Ghassemi, Marzyeh},
  journal={NPJ digital medicine},
  volume={4},
  number={1},
  pages={31},
  year={2021},
  publisher={Nature Publishing Group UK London}
}

@article{berteloot2026ai,
  title={AI-Driven Detection and Classification of Voice Disorders Using Acoustic Recordings},
  author={Berteloot, Louise and Sismono, Fergio and Maertens, L{\'e}onore and Maryn, Youri and Verguts, Monique and Vandaele, Melanie and Van de Pol, Charlotte and De Jaeger, Peter and Delsupehe, Kathelijne},
  journal={Journal of Voice},
  year={2026},
  doi={10.1016/j.jvoice.2026.04.024}
}

@article{barnett2024clinician,
  title={Improving clinician performance in classifying EEG patterns on the ictal--interictal injury continuum using interpretable machine learning},
  author={Barnett, Alina Jade and Guo, Zhicheng and Jing, Jin and Ge, Wendong and Kaplan, Peter W and Kong, Wan Yee and Karakis, Ioannis and Herlopian, Aline and Jayagopal, Lakshman Arcot and Taraschenko, Olga and others},
  journal={NEJM AI},
  volume={1},
  number={6},
  pages={AIoa2300331},
  year={2024},
  publisher={Massachusetts Medical Society},
  doi={10.1056/AIoa2300331}
}

@article{qazi2026automation,
  title={Automation bias in large language model--Assisted diagnostic reasoning among physicians trained in AI literacy---A randomized clinical trial},
  author={Qazi, Ihsan Ayyub and Ali, Ayesha and Khawaja, Asad Ullah and Akhtar, Muhammad Junaid and Sheikh, Ali Zafar and Alizai, Muhammad Hamad},
  journal={NEJM AI},
  volume={3},
  number={5},
  pages={AIoa2501001},
  year={2026},
  publisher={Massachusetts Medical Society},
  doi={10.1056/AIoa2501001}
}

@article{mairesse2026diagnostic,
  title={Diagnostic Accuracy of Artificial Intelligence in Laryngeal Disorders: An Integrative Review},
  author={Mairesse, Samantha and Maniaci, Antonino and Briganti, Giovanni and Lechien, Jerome R},
  journal={Journal of Personalized Medicine},
  volume={16},
  number={6},
  pages={301},
  year={2026},
  doi={10.3390/jpm16060301}
}

@article{goh2024llm,
  title={Large Language Model Influence on Diagnostic Reasoning: A Randomized Clinical Trial},
  author={Goh, Ethan and Gallo, Robert and Hom, Jason and Strong, Eric and Weng, Yingjie and Kerman, Hannah and Cool, Jos{\'e}phine A. and Kanjee, Zahir and Parsons, Andrew S. and Ahuja, Neera and others},
  journal={JAMA Network Open},
  volume={7},
  number={10},
  pages={e2440969},
  year={2024},
  doi={10.1001/jamanetworkopen.2024.40969}
}

@article{zwaan2026trust,
  title={Trust, Scrutiny, or Collaboration? A Performance-Based Framework for Human--AI Interaction in Medicine},
  author={Zwaan, Laura and Rodman, Adam and Shimizu, Taro},
  journal={NEJM AI},
  volume={3},
  number={5},
  pages={AIe2600354},
  year={2026},
  publisher={Massachusetts Medical Society},
  doi={10.1056/AIe2600354}
}

@article{https://doi.org/10.1002/lary.23630,
author = {Paul, Benjamin C. and Chen, Si and Sridharan, Shaum and Fang, Yixin and Amin, Milan R. and Branski, Ryan C.},
title = {Diagnostic accuracy of history, laryngoscopy, and stroboscopy},
journal = {The Laryngoscope},
volume = {123},
number = {1},
pages = {215-219},
year = {2013}
}

@article{dejonckere2001basic,
  title={A basic protocol for functional assessment of voice pathology, especially for investigating the efficacy of (phonosurgical) treatments and evaluating new assessment techniques: guideline elaborated by the Committee on Phoniatrics of the European Laryngological Society (ELS)},
  author={Dejonckere, Philippe H and Bradley, Patrick and Clemente, Pais and Cornut, Guy and Crevier-Buchman, Lise and Friedrich, Gerhard and Van De Heyning, Paul and Remacle, Marc and Woisard, Virginie},
  journal={European Archives of Oto-rhino-laryngology},
  volume={258},
  number={2},
  pages={77--82},
  year={2001},
  publisher={Springer}
}

@article{patel2018recommended,
  title={Recommended protocols for instrumental assessment of voice: American Speech-Language-Hearing Association expert panel to develop a protocol for instrumental assessment of vocal function},
  author={Patel, Rita R and Awan, Shaheen N and Barkmeier-Kraemer, Julie and Courey, Mark and Deliyski, Dimitar and Eadie, Tanya and Paul, Diane and {\v{S}}vec, Jan G and Hillman, Robert},
  journal={American journal of speech-language pathology},
  volume={27},
  number={3},
  pages={887--905},
  year={2018},
  publisher={American Speech-Language-Hearing Association}
}

@incollection{hirano1981psychoacoustic,
  author    = {Hirano, Minoru},
  title     = {Psycho-acoustic evaluation of voice},
  editor    = {Arnold, G. E. and Winckel, F. and Wyke, B. D.},
  booktitle = {Clinical Examination of Voice},
  volume    = {5},
  series    = {Disorders of Human Communication},
  pages     = {81--84},
  publisher = {Springer-Verlag},
  address   = {Vienna},
  year      = {1981}
}

@article{jacobson1997voice,
  title={The voice handicap index (VHI) development and validation},
  author={Jacobson, Barbara H and Johnson, Alex and Grywalski, Cynthia and Silbergleit, Alice and Jacobson, Gary and Benninger, Michael S and Newman, Craig W},
  journal={American journal of speech-language pathology},
  volume={6},
  number={3},
  pages={66--70},
  year={1997},
  publisher={American Speech-Language-Hearing Association}
}

@article{xu2010application,
  title={Application of the Mandarin Chinese version of the voice handicap index},
  author={Xu, Wen and Han, Demin and Li, Hongyan and Hu, Rong and Zhang, Li},
  journal={Journal of Voice},
  volume={24},
  number={6},
  pages={702--707},
  year={2010},
  publisher={Elsevier}
}

@article{speech_accessibility_project,
  title={Community-supported shared infrastructure in support of speech accessibility},
  author={Hasegawa-Johnson, Mark and Zheng, Xiuwen and Kim, Heejin and Mendes, Clarion and Dickinson, Meg and Hege, Erik and Zwilling, Chris and Channell, Marie Moore and Mattie, Laura and Hodges, Heather and others},
  journal={Journal of Speech, Language, and Hearing Research},
  volume={67},
  number={11},
  pages={4162--4175},
  year={2024},
  publisher={American Speech-Language-Hearing Association}
}

@incollection{qualispeech,
  title={{Q}uali{S}peech: A Speech Quality Assessment Dataset with Natural Language Reasoning and Descriptions},
  author={Wang, Siyin and Yu, Wenyi and Chen, Xianzhao and Tian, Xiaohai and Zhang, Jun and Lu, Lu and Tsao, Yu and Yamagishi, Junichi and Wang, Yuxuan and Zhang, Chao},
  editor={Che, Wanxiang and Nabende, Joyce and Shutova, Ekaterina and Pilehvar, Mohammad Taher},
  booktitle={Proceedings of the 63rd Annual Meeting of the Association for Computational Linguistics (Volume 1: Long Papers)},
  address={Vienna, Austria},
  publisher={Association for Computational Linguistics},
  pages={23588--23609},
  year={2025},
  doi={10.18653/v1/2025.acl-long.1150}
}

@article{svm,
  title={Support-vector networks},
  author={Cortes, Corinna and Vapnik, Vladimir},
  journal={Machine learning},
  volume={20},
  number={3},
  pages={273--297},
  year={1995},
  publisher={Springer}
}

@article{gmm_hmm,
  title={A tutorial on hidden Markov models and selected applications in speech recognition},
  author={Rabiner, Lawrence R},
  journal={Proceedings of the IEEE},
  volume={77},
  number={2},
  pages={257--286},
  year={1989},
  publisher={Ieee}
}

@article{mesallam2017development,
  title={Development of the arabic voice pathology database and its evaluation by using speech features and machine learning algorithms},
  author={Mesallam, Tamer A and Farahat, Mohamed and Malki, Khalid H and Alsulaiman, Mansour and Ali, Zulfiqar and Al-Nasheri, Ahmed and Muhammad, Ghulam},
  journal={Journal of healthcare engineering},
  volume={2017},
  number={1},
  pages={8783751},
  year={2017},
  publisher={Wiley Online Library}
}

@article{gmm_i_vector,
  title={Front-end factor analysis for speaker verification},
  author={Dehak, Najim and Kenny, Patrick J and Dehak, R{\'e}da and Dumouchel, Pierre and Ouellet, Pierre},
  journal={IEEE Transactions on Audio, Speech, and Language Processing},
  volume={19},
  number={4},
  pages={788--798},
  year={2011},
  publisher={IEEE}
}

@article{lad2019total,
  title={Total variability factor analysis for dysphonia detection},
  author={Lad, Nikunj R and Nirmal, JH and Naikare, Kshipra D},
  journal={International Journal of Information Technology},
  volume={11},
  number={1},
  pages={67--74},
  year={2019},
  publisher={Springer}
}

@article{wav2vec2,
  title={wav2vec 2.0: A framework for self-supervised learning of speech representations},
  author={Baevski, Alexei and Zhou, Yuhao and Mohamed, Abdelrahman and Auli, Michael},
  journal={Advances in neural information processing systems},
  volume={33},
  pages={12449--12460},
  year={2020}
}

@article{cai2024voice,
  title={Voice disorder classification using wav2vec 2.0 feature extraction},
  author={Cai, Jie and Song, Yuliang and Wu, Jianghao and Chen, Xiong},
  journal={Journal of Voice},
  year={2024},
  publisher={Elsevier}
}

@article{tirronen2023hierarchical,
  title={Hierarchical multi-class classification of voice disorders using self-supervised models and glottal features},
  author={Tirronen, Saska and Kadiri, Sudarsana Reddy and Alku, Paavo},
  journal={IEEE Open Journal of Signal Processing},
  volume={4},
  pages={80--88},
  year={2023},
  publisher={IEEE}
}

@inproceedings{whisper,
  title     = {Robust Speech Recognition via Large-Scale Weak Supervision},
  author    = {Radford, Alec and Kim, Jong Wook and Xu, Tao and Brockman, Greg and McLeavey, Christine and Sutskever, Ilya},
  editor    = {Krause, Andreas and Brunskill, Emma and Cho, Kyunghyun and Engelhardt, Barbara and Sabato, Sivan and Scarlett, Jonathan},
  booktitle = {Proceedings of the 40th International Conference on Machine Learning},
  series    = {Proceedings of Machine Learning Research},
  volume    = {202},
  pages     = {28492--28518},
  year      = {2023},
  publisher = {PMLR}
}

@article{merler2025clinical,
  title={Clinical assessment and interpretation of dysarthria in ALS using attention based deep learning AI models},
  author={Merler, Michele and Agurto, Carla and Peller, Julian and Roitberg, Esteban and Taitz, Alan and Trevisan, Marcos A and Navar, Indu and Berry, James D and Fraenkel, Ernest and Ostrow, Lyle W and others},
  journal={NPJ Digital Medicine},
  volume={8},
  number={1},
  pages={260},
  year={2025},
  publisher={Nature Publishing Group UK London}
}

@inproceedings{lora,
  title     = {{LoRA}: Low-Rank Adaptation of Large Language Models},
  author    = {Hu, Edward J. and Shen, Yelong and Wallis, Phillip and Allen-Zhu, Zeyuan and Li, Yuanzhi and Wang, Shean and Wang, Lu and Chen, Weizhu},
  editor    = {Liu, Yan and Finn, Chelsea and Choi, Yejin and Deisenroth, Marc},
  booktitle = {International Conference on Learning Representations},
  year      = {2022},
}

@article{koo2016guideline,
  title={A guideline of selecting and reporting intraclass correlation coefficients for reliability research},
  author={Koo, Terry K and Li, Mae Y},
  journal={Journal of chiropractic medicine},
  volume={15},
  number={2},
  pages={155--163},
  year={2016},
  publisher={National University of Health Sciences}
}

@article{jabbour2023,
  author  = {Jabbour, Sarah and Fouhey, David and Shepard, Stephanie and
             Valley, Thomas S. and Kazerooni, Ella A. and Banovic, Nikola and
             Wiens, Jenna and Sjoding, Michael W.},
  title   = {Measuring the impact of {AI} in the diagnosis of hospitalized
             patients: a randomized clinical vignette survey study},
  journal = {JAMA},
  year    = {2023},
  volume  = {330},
  number  = {23},
  pages   = {2275--2284},
  doi     = {10.1001/jama.2023.22295}
}

@article{chanda2024,
  author  = {Chanda, Tirtha and Hauser, Katja and Hobelsberger, Sarah and
             Bucher, Tabea-Clara and Garcia, Carina Nogueira and Wies, Christoph
             and others},
  title   = {Dermatologist-like explainable {AI} enhances trust and confidence
             in diagnosing melanoma},
  journal = {Nature Communications},
  year    = {2024},
  volume  = {15},
  pages   = {524},
  doi     = {10.1038/s41467-023-43095-4}
}

\setcounter{section}{0}
\renewcommand{\thesection}{S\arabic{section}}
\renewcommand{\theHsection}{S\arabic{section}}
\setcounter{table}{0}
\renewcommand{\thetable}{S\arabic{table}}
\renewcommand{\theHtable}{S\arabic{table}}
\setcounter{figure}{0}
\renewcommand{\thefigure}{S\arabic{figure}}
\renewcommand{\theHfigure}{S\arabic{figure}}
\setcounter{equation}{0}
\renewcommand{\theequation}{S\arabic{equation}}
\renewcommand{\theHequation}{S\arabic{equation}}

\clearpage
\section*{Supplementary Information}
\label{sec:supp}

\renewcommand{\topfraction}{0.95}
\renewcommand{\bottomfraction}{0.95}
\renewcommand{\textfraction}{0.05}
\renewcommand{\floatpagefraction}{0.8}

\section{Supplementary Methods}
\label{sec:supp_methods}

\subsection{Laryngeal Voice-Recording Protocol}
\label{sec:supp_methods_recording}

The standardized laryngeal voice assessment comprised sustained phonation of /a/ and /i/ at a comfortable pitch and loudness and reading of a standardized Mandarin sentence. Recordings were acquired in a quiet clinical room at a sampling rate of at least 44.1 kHz with 16-bit resolution, using a head-mounted or table-top condenser microphone. Healthy controls were recorded in the same room and with the same equipment as participants with laryngeal disease. Audio was resampled to 16 kHz before model training and inference, and recording-device identifiers were not provided to the model.

\subsection{Model Development, Training Sources, and Multi-task Fine-tuning}
\label{sec:supp_methods_training}

\subsubsection{Base model and parameter-efficient adaptation}

The model was Qwen3-Omni-30B-A3B-Thinking \cite{qwen2025omni}, adapted with low-rank adaptation (LoRA) \cite{lora} on all linear modules (rank 8, scaling factor 32, dropout 0.05). Training used fused AdamW (weight decay 0.1, gradient clipping at norm 1.0), a learning rate of $1\times10^{-5}$ with a cosine schedule and 0.05 warm-up ratio, bfloat16 mixed precision, 2048-token sequences, and a per-device batch size of 32 with one gradient-accumulation step (effective global batch 256 on 8 GPUs in Stage 1 and 512 on 16 GPUs in Stage 2), on NVIDIA H200 GPUs with the ms-swift framework and a global random seed of 42. Patient-level partitions were fixed before audio-view construction, keeping each participant's recordings in a single partition.

\subsubsection{Training sources and task balancing}

Training combined laryngeal tasks (seven-class provisional categorization and perceptual voice-phenotype tasks constructed from the clinical cohort described in the main Methods), neurological dysarthria data from the Speech Accessibility Project \cite{speech_accessibility_project}, and general auditory-assessment supervision from QualiSpeech \cite{qualispeech}; none of the additional domains contributed to the held-out laryngeal test cohort or the reader study. Minority categories were oversampled within each clinical training partition, with ordinal or continuous targets first binned, leaving validation and held-out distributions unchanged; clinical source $i$ with $n_i$ eligible examples was then sampled with probability $p_i = n_i^{0.4}/\sum_j n_j^{0.4}$ to limit domination by the largest sources. The unified model was compared with task-specific models on each task family (Tables~\ref{tab:laryngeal_etiology_baseline_comparison} and \ref{tab:etiology_prediction}).

\subsubsection{Secondary dysarthria cohort and tasks}

The English dysarthria data were drawn from the Speech Accessibility Project \cite{speech_accessibility_project}: 456 participants (11,908 recordings) with perceptual severity ratings on a 1-to-7 scale and one of four etiologies (Parkinson's disease, cerebral palsy, amyotrophic lateral sclerosis, or Down syndrome), partitioned into 373 training, 32 validation, and 51 held-out participants.
Additional etiology-labeled recordings from the same source, without perceptual severity ratings, were used for training only. After retaining rationale targets only when the predicted etiology agreed with the reference label, 194,423 training examples remained.
Task-specific auditory large language models and the comparator systems of Section \ref{sec:supp_methods_comparators} were trained for each task on the same partitions. 
Perceptual severity was evaluated at the recording level using seven-class and collapsed four-class accuracy, mean absolute error, and Pearson and Spearman correlation coefficients. Etiologic categorization was evaluated at the patient level using weighted precision, recall, and F1, with patient-level predictions obtained by majority vote across recordings.

\subsubsection{Two-stage unified multi-task fine-tuning and checkpoint selection}

Fine-tuning ran in two one-epoch stages. Stage 1 emphasized general auditory assessment (predominantly QualiSpeech); its 800-step checkpoint initialized Stage 2, which emphasized clinical voice assessment and categorization, with 10\% QualiSpeech rehearsal and about four effective passes over the pooled clinical sources. The Stage-2 checkpoint at 8000 steps was selected on validation patient-level weighted F1 for seven-class laryngeal categorization and was fixed before held-out testing and the reader study.

\subsection{Laryngeal Instruction Construction and Rationale Supervision}
\label{sec:supp_methods_laryngeal_supervision}

Each laryngeal training example paired an audio input, the phonation or speech-task information, a task instruction, and the pre-laryngoscopic inputs defined in the main Methods (voice use was recorded as daily duration and intensity and years of work experience). For seven-class categorization, the instruction asked the model to assess audible characteristics (overall severity, roughness, breathiness, asthenia, strain, pitch abnormality, and phonatory instability), integrate them with the pre-laryngoscopic information, and assign one category; the target response contained the category and an acoustic rationale linking it to the perceptual and acoustic phenotypes. Rationale targets were constructed from the pre-laryngoscopic inputs, the reference GRBAS ratings, phenotype rules, and the adjudicated category; rule-based checks verified their consistency with the reference label and with the evidence fields, and the templates and representative examples were reviewed by voice clinicians before fine-tuning (Supplementary Table~\ref{tab:supp_prompt_templates}). Alternative rationale configurations evaluated during development are described in Section~\ref{sec:supp_methods_robustness}.

Ten auxiliary tasks, the five GRBAS dimensions, pitch status, and VHI-30 total and functional, physical, and emotional subscale scores \cite{jacobson1997voice,xu2010application}, used dimension-specific instructions with the same input format. Reference categories, GRBAS ratings, and VHI-30 scores served only as supervision targets: they were not model inputs at validation or held-out inference and were not shown to readers.

\subsection{Audio-View Construction, Inference Procedures, and Output Processing}
\label{sec:supp_methods_inference}

\subsubsection{Audio-view construction}

An original utterance was one separately acquired recording, and an audio view was either one original utterance or a within-patient concatenation of two or more utterances from the same assessment. Each audio view corresponds to one reading of the model as described in the main text. For a participant with $n$ utterances, every nonempty subset formed one view, giving up to $2^n-1$ views (three utterances gave three single-utterance, three two-utterance, and one three-utterance view); utterance order within a concatenation was permuted with a fixed seed of 42. During development the derived views served as within-patient augmentation rather than as independent observations, and the same construction was used for held-out inference, without duration-based truncation; the longest view was 34.5 seconds. In the held-out cohort, the number of valid views per participant ranged from 1 to 63 (median, 7).

\subsubsection{Decoding and audio-view outputs}

Held-out inference used deterministic greedy decoding with the ms-swift PyTorch inference backend. The maximum generation length was 4096 new tokens and the maximum model context length was 8192 tokens. Generation terminated when an end-of-sequence token was emitted or when the maximum generation length was reached. The same decoding settings were used for standalone held-out evaluation and generation of the reader-study assessment cards.

\subsubsection{Category parsing and patient-level aggregation}

A deterministic parser extracted the final category field from each generated response and mapped it to one of the seven canonical laryngeal categories. Fixed answer prefixes, surrounding quotation marks, terminal punctuation, and extraneous whitespace were removed before category matching.
A normalization dictionary mapped accepted synonymous or finer-grained diagnostic expressions to the corresponding broad category. For example, vocal fold nodules, cysts, polyps, and Reinke edema were mapped to the benign vocal fold lesion category.
An output was considered valid only when it could be mapped unambiguously to exactly one canonical category. Outputs that could not be mapped to one of the seven categories were excluded from categorical aggregation.

The final patient-level category was determined by majority voting across all valid audio-view predictions. If multiple categories shared the highest vote count, one of the tied categories was selected using a random tie-resolution procedure with seed 42. The resulting patient-level category was used consistently for standalone metric calculation, reader--model concordance, and assessment-card display. Five participants in the held-out test cohort had tied highest vote counts. No held-out participant lacked a valid category prediction across all available views.

Within-patient prediction consistency was calculated as

\begin{equation}
C_i =
\frac{N_{i,\mathrm{majority}}}
{N_{i,\mathrm{valid}}},
\end{equation}

where $N_{i,\mathrm{majority}}$ denotes the number of valid audio-view predictions assigned to the final patient-level category and $N_{i,\mathrm{valid}}$ denotes the total number of valid category predictions for participant $i$. 

The category vote distribution recorded the number of valid audio-view predictions assigned to each of the seven categories.

\subsubsection{Malignancy alert}
\label{sec:supp_methods_alert}

The malignancy alert was defined on the vote fraction, the proportion of valid audio-view predictions assigned to precancerous or malignant lesions. The alert threshold was the highest vote fraction at which sensitivity in the validation set remained at least 80\%, which was 2/7 (0.2857); the alert was raised when the vote fraction was at least 2/7. At this threshold, sensitivity was 86.1\% and specificity was 78.6\% in the validation set. The threshold was fixed before held-out testing and applied unchanged in the temporal-validation cohort.

\subsubsection{Patient-level voice-assessment aggregation and rationale selection}

GRBAS, VHI-30, and pitch outcomes were evaluated at the utterance and patient levels. Patient-level GRBAS and VHI-30 estimates and continuous pitch values were calculated as the means across all valid audio-view predictions for each participant. Patient-level categorical pitch status was determined by majority vote across valid audio-view predictions; ties were resolved according to the direction of the mean signed pitch estimate.

The displayed acoustic rationale was selected only after the final patient-level category had been determined. Candidate audio-view outputs were searched in descending order of the number of constituent original utterances, and the first output whose category prediction agreed with the final patient-level category was used as the displayed rationale. This rule prioritized a rationale derived from the broadest available within-patient acoustic sample while maintaining agreement between the displayed rationale and the final category.

\subsubsection{Temporal-validation view construction and inference} 
\label{sec:temp-supp}

Views were constructed as in the development cohort for participants with seven or fewer original utterances. For participants with more, the number of views was capped at 128: all single-utterance views and the concatenation of all utterances were retained, and the remaining budget was filled with combinations of intermediate length, allocated in proportion to the number available at each length and sampled with a patient-specific fixed seed. The cap depended only on recording availability and was defined independently of reference labels and model predictions.

\subsection{Standalone Comparator Models}
\label{sec:supp_methods_comparators}

All comparators used the same patient-level partitions as the auditory large language model, were optimized separately for each task, selected on validation performance, and evaluated with the same measures and patient-level aggregation as the corresponding model analysis. Configurations for seven-class laryngeal categorization are given in Supplementary Table~\ref{tab:supp_baseline_configurations}.

\subsubsection{Classical acoustic models}

Hand-crafted acoustic features comprised duration, energy, spectral, mel-frequency cepstral (with temporal derivatives), pitch, perturbation, and harmonics-to-noise descriptors. Three families were evaluated: support-vector machines \cite{svm,kim2024classification} on the pooled features; a GMM-i-vector system \cite{gmm_i_vector,lad2019total}, in which a Gaussian-mixture universal background model and total-variability representation fed a support-vector predictor; and class-conditional GMM-HMMs \cite{gmm_hmm,mesallam2017development}, which assigned the category with the highest likelihood. For laryngeal categorization, the SVM and GMM-i-vector comparators also received age, body-mass index, sex, fundamental frequency, jitter, shimmer, noise-to-harmonics ratio, and maximum phonation time as covariates, concatenated with the i-vector in the latter; the GMM-HMM used acoustic features alone. 
For the dysarthria tasks, harmonized non-audio covariates were unavailable, and the classical comparators therefore used acoustic features alone.

\subsubsection{Pretrained audio encoders}

Wav2Vec2-Large \cite{wav2vec2,cai2024voice,tirronen2023hierarchical} and Whisper-large-v3 \cite{whisper,merler2025clinical} were fine-tuned with LoRA at the same rank, scaling factor, and dropout as the auditory large language model, and their pooled representations fed task-specific linear heads for classification, regression, or multilabel prediction. A task-specific auditory large language model trained for the same endpoint was also evaluated alongside the unified model.

\subsection{Standalone Robustness and Development-Stage Configuration Analyses}
\label{sec:supp_methods_robustness}

\subsubsection{Robustness to audio-view selection}

Robustness of the model to the choice of patient-level audio representation was evaluated using three view sets. The original-utterance-only analysis aggregated predictions from the independently acquired original utterances without derived concatenations. The single-merged-view analysis used one concatenated view containing all available original utterances from each participant. The primary all-view analysis used the complete set of original and within-patient concatenated audio views described in Supplementary Methods Section~\ref{sec:supp_methods_inference}.

Seven-class patient-level accuracy and weighted F1 were recalculated for each view set. 
Sensitivity and specificity for precancerous or malignant lesions under the seven-class prediction were also recalculated to assess the dependence of high-risk recognition on the audio-view aggregation scheme. 
Detailed results are reported in Supplementary Table~\ref{tab:supp_aggregation_sensitivity}.

\subsubsection{Tie-resolution sensitivity}

Five held-out participants had tied highest vote counts under the primary all-view aggregation procedure. The primary analysis retained the seed-controlled tie-resolution rule described in Supplementary Methods Section~\ref{sec:supp_methods_inference}. To assess the influence of these ties, patient-level metrics were recomputed across all permissible assignments of the tied cases while leaving all non-tied predictions unchanged.

For each metric, the minimum and maximum values attainable across the permissible tie resolutions were determined. The analysis included seven-class accuracy and weighted F1, together with sensitivity and specificity for precancerous or malignant lesions. Results are reported in Supplementary Table~\ref{tab:supp_tie_sensitivity}.

\subsubsection{Development-stage task-specific configuration analyses}

Separate task-specific development-stage analyses were conducted for seven-class laryngeal categorization to examine the effects of audio-view construction, available model inputs, and rationale supervision. 
These analyses used task-specific laryngeal models trained specifically for the seven-class categorization task and were distinct from the unified model evaluated in the primary standalone analyses and reader study.

During task-specific model development, the same three audio-view configurations defined above were compared within the rule-constrained rationale-supervision series to assess the effect of view construction on seven-class laryngeal categorization.
The all-view configuration was then used to examine the contribution of structured pre-visualization inputs. The no-objective-voice-measures configuration omitted fundamental frequency, jitter, shimmer, noise-to-harmonics ratio, and maximum phonation time while retaining audio, task information, and available patient context. The no-patient-context configuration omitted demographic, occupational, and voice-use information while retaining audio, task information, and the objective voice measures.

Rationale supervision was evaluated under the full-input, all-view setting using three configurations that shared the same model inputs and final category target but differed in the assistant-side rationale: rule-constrained, in which the rationale followed structured templates centered on perceptual voice attributes; free-text, in which it was expressed in less templated clinical prose that integrated the available supervisory evidence more flexibly; and empty-rationale, which omitted rationale text. 
The unified model used for the primary analyses was trained separately as described in Supplementary Methods Section~\ref{sec:supp_methods_training}.

For each development-stage configuration, patient-level weighted precision, weighted recall, and weighted F1 were evaluated. Categorical predictions used the aggregation and tie-resolution procedures applicable to the corresponding audio-view setting.

Detailed results are provided in Supplementary Table~\ref{tab:supp_configuration_analysis}.

\subsection{Reader-Study Materials and Implementation}
\label{sec:supp_methods_reader}

Each of the 32 readers was independently assigned 25 participants sampled at random without replacement from the 195-participant held-out test set. Participants could be assigned to more than one reader; the assignments yielded 800 paired reader--case assessments involving 192 unique participants. The resulting distribution across reference categories is reported in Table~\ref{tab:reader_disease_specific}. Reader characteristics, including the experience-group definition used in subgroup analyses, are given in Table~\ref{tab:reader_characteristics}.

\subsection{Statistical Implementation and Reproducibility}
\label{sec:supp_methods_statistics}

Unless otherwise specified, 95\% confidence intervals were percentile intervals from 5000 bootstrap resamples. Standalone measures, including the malignancy-alert measures, AUROC, and GRBAS agreement, were resampled at the participant level. Reader-study measures (between-round changes in accuracy, macro-recall, macro-F1, and sensitivity and specificity for precancerous or malignant lesions, and the ratio of incorrect-to-correct to correct-to-incorrect transitions) used a two-way clustered bootstrap: readers and participants were resampled independently, each reader--participant observation was weighted by the product of its reader and participant multiplicities, and paired unaided and AI-assisted responses were kept together.

GRBAS agreement was quantified with a two-way random-effects, absolute-agreement, single-measure intraclass correlation coefficient (ICC[2,1]) \cite{koo2016guideline} for each dimension, together with Pearson and Spearman correlation coefficients; the across-dimension mean was descriptive.

The primary reader-study model was fitted with \texttt{lme4::glmer} (binomial logit, \texttt{bobyqa} optimizer) with random intercepts for reader and participant, and the effect of study round was summarized as an odds ratio with a Wald 95\% confidence interval and a two-sided Wald $P$ value. Two sensitivity models added a random intercept for each reader--participant dyad and, in the second model, also a reader-specific random slope for study round (Table~\ref{tab:reader_primary_robustness}).

Exploratory heterogeneity was evaluated by adding each factor and its interaction with study round to the primary model: reader experience group, professional level (as a four-level factor and as an ordered variable from postgraduate clinical trainee to chief physician), practice region, reference category, within-patient prediction consistency, and whether the model was correct; factors with more than two levels were tested with a global likelihood-ratio test. Heterogeneity in reader--model concordance, and in response matching (selection of the model category in the AI-assisted round after an unaided response that differed from it), was examined according to experience group, professional level, and practice region; because these are reader-level characteristics, models allowing reader-specific study-round effects and reader-level permutation analyses treating the 32 readers as the independent units were also used. Holm-adjusted $P$ values were calculated separately within the three exploratory families (reader-performance heterogeneity, concordance heterogeneity, and response matching).

Statistical analyses were performed using Python 3.10.12 with scikit-learn 1.7.2 and R 4.5.3 with lme4 2.0-6.

\section{Supplementary Results}
\label{sec:supp_results}

\subsection{Temporal Validation} 
\label{sec:supp_results_temporal} 

Application of the view-construction procedure to the 605 recordings of the temporal-validation cohort yielded 2199 audio views, and patient-level predictions were obtained with the same parser, aggregation, and tie-resolution rules as in the held-out evaluation. Class-specific performance is shown in Table~\ref{tab:temporal_validation_class_performance}. 

\subsection{Additional Standalone Evaluation of the Model}
\label{sec:supp_results_standalone}

\subsubsection{Comparator performance}

Patient-level results for all comparator systems are given in Table~\ref{tab:laryngeal_etiology_baseline_comparison}.

All 20 healthy controls were classified as healthy, and none of the 175 participants with a disease category was classified as healthy (Fig.~\ref{fig:2}A), so that healthy control versus disorder was separated with 100\% accuracy in this cohort. Because the controls had no voice complaints and were not drawn from patients presenting with hoarseness, this result is not an estimate of screening performance.

\subsubsection{Robustness to audio-view selection and tie resolution}

The primary all-view procedure gave higher weighted F1 and high-risk sensitivity than aggregation over original utterances only or over a single merged view, with similar specificity (Table~\ref{tab:supp_aggregation_sensitivity}). Across all permissible resolutions of the five tied votes, weighted F1 ranged from 0.71 to 0.73 and high-risk sensitivity from 71.4\% to 77.1\% (Table~\ref{tab:supp_tie_sensitivity}); the minimum values remained above those of the alternative view schemes.

\subsubsection{Development-stage task-specific configuration analyses}

Among the task-specific development-stage configurations (Table~\ref{tab:supp_configuration_analysis}), aggregation over all views outperformed original utterances only or a single merged view; omitting the acoustic measures or the patient-context fields reduced patient-level weighted F1 by 0.02 and 0.03, respectively; and free-text rationale supervision outperformed rule-constrained and empty-rationale supervision (weighted F1, 0.68, 0.65, and 0.59). All evaluated configurations retained the audio views, so these analyses did not isolate the contribution of raw audio relative to patient context and routine acoustic measures alone.

\subsection{Additional Voice-Assessment Outcomes}
\label{sec:supp_results_voice}

\subsubsection{GRBAS agreement with senior-reviewed ratings}

Agreement with the reference GRBAS ratings was higher at the patient level than at the utterance level (overall mean absolute error, 0.25 vs. 0.33; intraclass correlation coefficient, 0.73 vs. 0.59). Dimension-specific values and comparator results are given in Tables \ref{tab:grbas_mae} to \ref{tab:grbas_spearman}.

\subsubsection{Auxiliary VHI-30 and pitch outcomes}

At the patient level, intraclass correlation coefficients were 0.53 for the VHI-30 total score (164 participants with reference scores) and 0.63 for continuous pitch, compared with 0.73 across the GRBAS dimensions (Tables \ref{tab:vhi_mae} to \ref{tab:pitch_results}). Neither VHI-30 nor pitch was displayed on the assessment card.

\subsection{Exploratory Reader-Study Analyses}
\label{sec:supp_results_reader}

\subsubsection{Reader subgroup, disease-specific, and prediction-consistency analyses}

Across the four professional levels, the between-round difference in categorization accuracy ranged from 13.3 percentage points among postgraduate clinical trainees to 3.3 percentage points among chief physicians. An exploratory ordinal analysis yielded an interaction odds ratio of 0.88 per one-level increase in professional level (95\% CI, 0.69 to 1.11). Interaction analyses did not provide clear evidence that the between-round difference varied according to experience group or professional level (Supplementary Table~\ref{tab:reader_exploratory_interactions}).

Across the seven reference categories, F1 was numerically higher in the AI-assisted round for every category. Recall and precision were each numerically higher for six of the seven categories, whereas specificity was higher for five categories and lower for two (Table~\ref{tab:reader_disease_specific}).

\subsubsection{Reader performance according to whether the model was correct}

The balance of paired correctness transitions differed according to whether the model was correct. When the model was correct, 94 of 536 assessments (17.5\%) changed from incorrect to correct and 15 (2.8\%) changed from correct to incorrect. When the model was incorrect, 5 of 264 assessments (1.9\%) changed from incorrect to correct and 28 (10.6\%) changed from correct to incorrect. Stratum-specific performance and complete transition distributions are provided in Supplementary Tables~\ref{tab:reader_model_correctness} and~\ref{tab:reader_model_correctness_transitions}.

\subsubsection{Reader--model concordance and responses after initial disagreement}

Reader--model concordance increased from 46.4\% in the unaided round to 63.0\% in the subsequent AI-assisted round. Among less-experienced readers, concordance increased from 44.8\% to 65.8\% (21.0 percentage points), compared with an increase from 48.0\% to 60.3\% (12.3 percentage points) among more-experienced readers, and the between-round increase ranged from 34.7 percentage points among postgraduate clinical trainees to 4.7 percentage points among chief physicians. Reader--case-level mixed-effects analyses suggested heterogeneity according to experience group and professional level, but these patterns were not reproduced in random-reader-slope models or reader-level permutation analyses (Supplementary Table~\ref{tab:reader_exploratory_interactions}).

Among the 429 reader--case assessments in which the unaided reader response differed from the model category, 151 (35.2\%) had a subsequent AI-assisted response that matched the model category. Response matching was more frequent among less-experienced than more-experienced readers (44.8\% vs. 25.0\%) and ranged from 67.4\% among postgraduate clinical trainees to 9.2\% among chief physicians, although the exploratory interaction analyses did not provide clear evidence of heterogeneity after within-family Holm adjustment.


\subsection{Performance on Secondary Dysarthria Tasks}
\label{sec:supp_results_secondary}

\begin{figure}[!htbp]
\centering
\includegraphics[width=\linewidth]{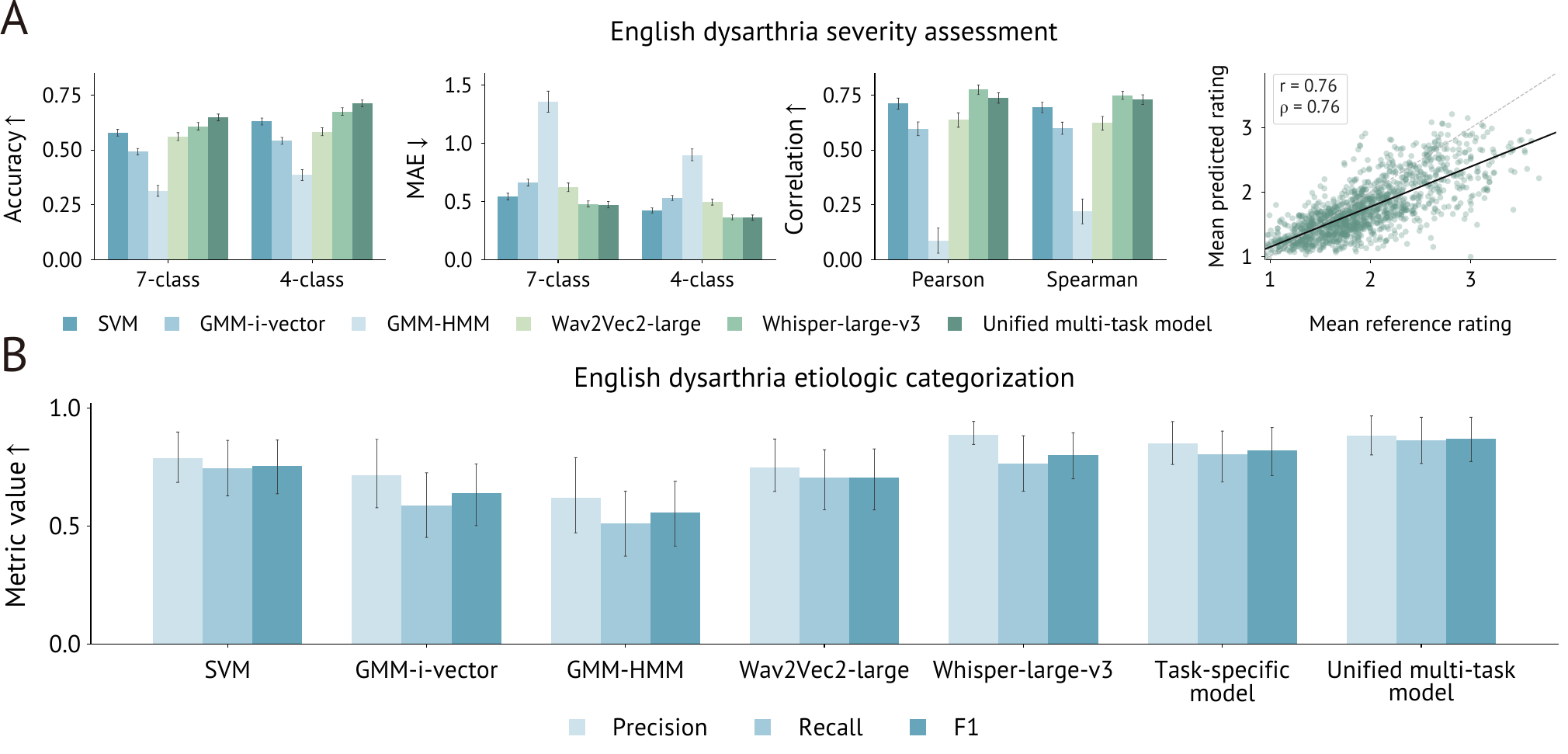}
\caption{Secondary held-out performance in the English dysarthria cohort.
\textbf{a}, Performance on English dysarthria perceptual-severity assessment, including seven- and four-class rating accuracy, mean absolute error (MAE), Pearson and Spearman correlations, and audio-recording-level agreement between predicted and reference severity ratings.
In the scatter plot, each point represents one audio recording, with its coordinates defined by the averaged predicted and reference severity ratings. The dashed line denotes identity, and the solid line denotes the fitted linear trend.
\textbf{b}, Patient-level performance for four-class English dysarthria etiologic categorization across conventional machine-learning baselines, pretrained speech baselines, the task-specific and unified multi-task models.
Bars show weighted precision, weighted recall and weighted F1.
Where shown, error bars denote 95\% bootstrap confidence intervals.
}
\label{fig:supp_dysarthria}
\end{figure}

\begin{figure}[htbp]
\centering
\includegraphics[width=\linewidth]{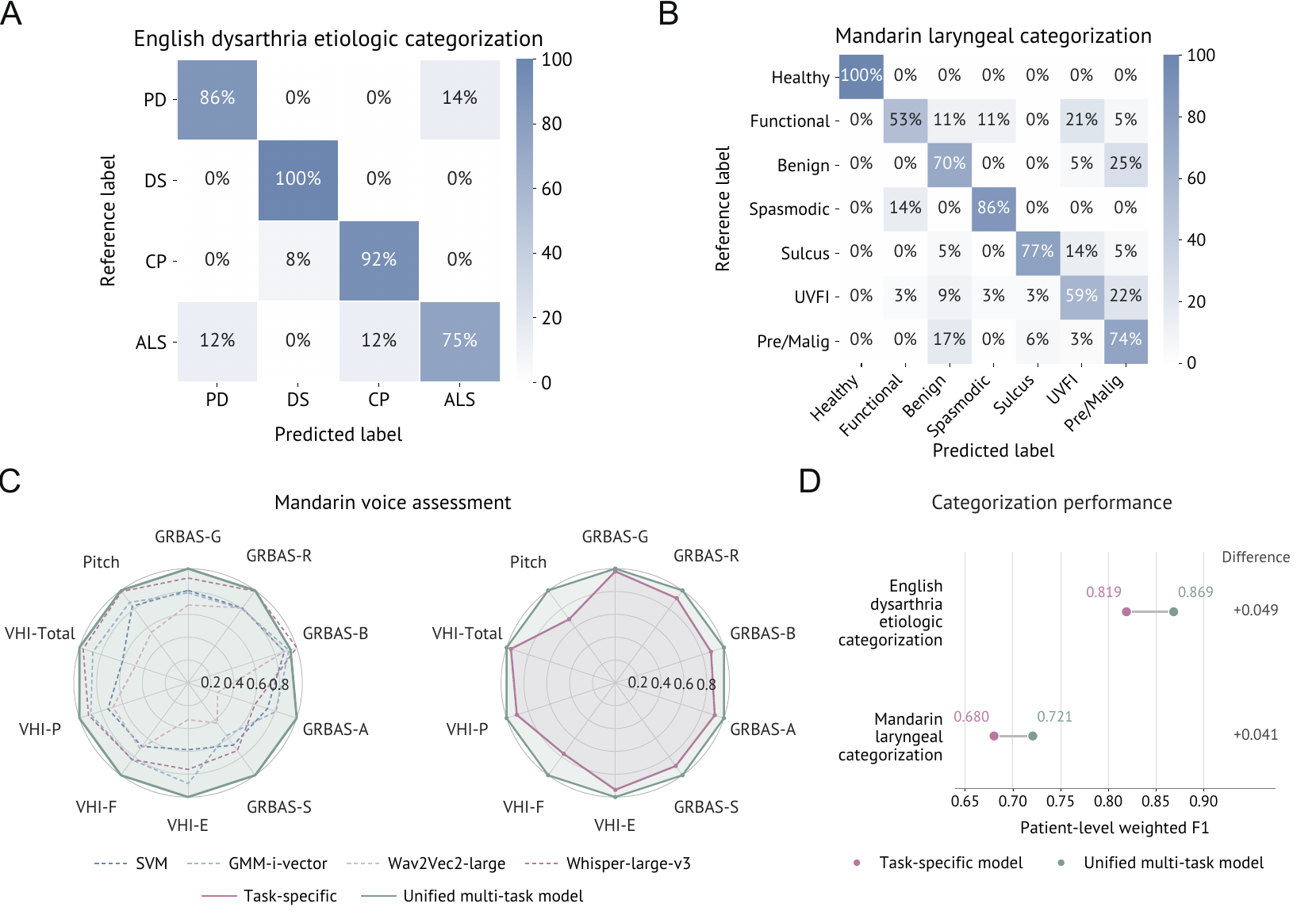}
\caption{Additional held-out performance summaries for the laryngeal and dysarthria tasks.
\textbf{a}, Patient-level row-normalized confusion matrix for four-class English dysarthria etiologic categorization by the unified multi-task model.
\textbf{b}, Patient-level row-normalized confusion matrix for seven-class provisional Mandarin laryngeal categorization by the unified multi-task model.
\textbf{c}, Summary of held-out patient-level intraclass correlation coefficients (ICCs) across Mandarin GRBAS, Voice Handicap Index, and pitch-assessment endpoints for the evaluated comparators (left) and the task-specific and unified multi-task models (right), with values for each endpoint normalized to the highest value among the models shown in the corresponding radar plot.
\textbf{d}, Patient-level weighted F1 for English dysarthria etiologic categorization and Mandarin laryngeal categorization with the task-specific and unified multi-task models. Differences denote the unified multi-task model value minus the task-specific model value.
PD denotes Parkinson's disease; DS, Down syndrome; CP, cerebral palsy; ALS, amyotrophic lateral sclerosis; UVFI, unilateral vocal fold immobility; and Pre/Malig, precancerous or malignant lesions.
}
\label{fig:supp_multitask_wo_pumch}
\end{figure}

Secondary performance was evaluated in the held-out dysarthria cohort.
For perceptual-severity assessment, predicted and reference ratings had Pearson and Spearman correlations of 0.76 (Fig.~\ref{fig:supp_dysarthria}A).
For etiologic categorization, patient-level weighted F1 was 0.87 with the unified model and 0.82 with the task-specific model, indicating that joint training did not reduce performance on this secondary task (Table~\ref{tab:etiology_prediction}, Fig.~\ref{fig:supp_dysarthria}B, and Fig.~\ref{fig:supp_multitask_wo_pumch}D).

\clearpage
\section{Supplementary Figures}
\label{sec:supp_figures}

\begin{figure}[!htbp]
\centering
\includegraphics[width=\linewidth]{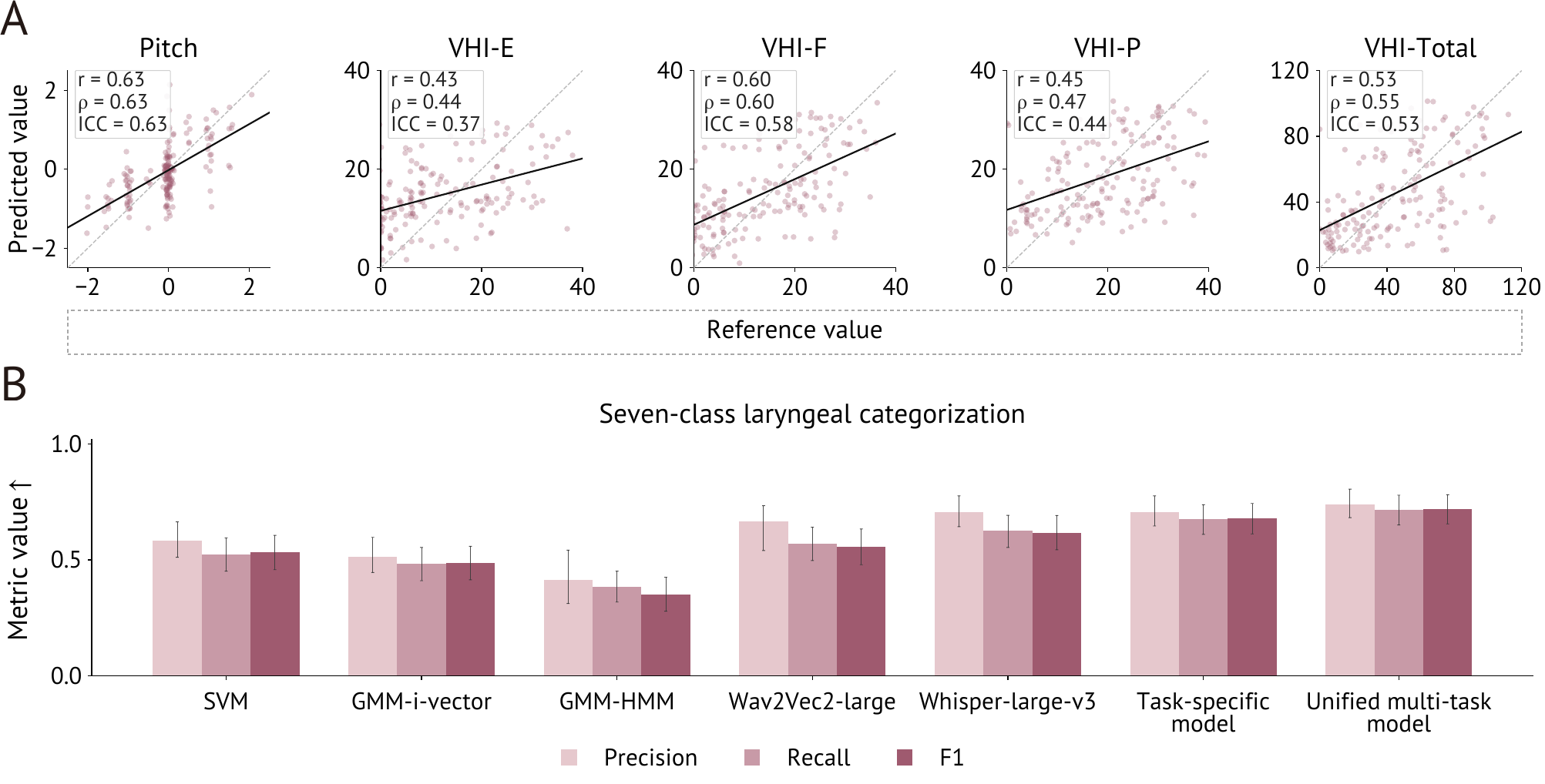}
\caption{Additional standalone performance in the held-out laryngeal test set.
\textbf{a}, Patient-level agreement between model-generated and reference continuous pitch values and Voice Handicap Index-30 (VHI-30) emotional (VHI-E), functional (VHI-F), physical (VHI-P), and total scores.
\textbf{b}, Patient-level performance for seven-class laryngeal categorization across conventional machine-learning comparators, pretrained speech comparators, the task-specific auditory large language model, and the unified multi-task model.
Bars show weighted precision, weighted recall, and weighted F1; error bars denote 95\% bootstrap confidence intervals.
}
\label{fig:supp_laryngeal_standalone}
\end{figure}

\clearpage
\section{Supplementary Tables}
\label{sec:supp_tables}

\begin{table}[!htbp]
\centering
\caption{Representative training formulations for rationale supervision in seven-class laryngeal categorization.}
\label{tab:supp_prompt_templates}
\footnotesize
\setlength{\tabcolsep}{3pt}
\renewcommand{\arraystretch}{1.05}
\begin{tabular}{
p{0.21\textwidth}
p{0.14\textwidth}
p{0.59\textwidth}
}
\toprule
\textbf{Configuration} & \textbf{Component} & \textbf{Representative training example} \\
\midrule
\textbf{Rule-constrained rationale} 
& \textit{Instruction} &
Listen to the recording from the perspective of clinical voice assessment and evaluate the degree of voice abnormality, breath control, pitch use, and phonatory strain. Integrate the audible characteristics with the available objective voice measures and patient information to infer the most likely laryngeal category. \\
& \begin{tabular}[t]{@{}l@{}} \textit{Rationale}\\ \textit{target}
\end{tabular} &
Standardized perceptual assessment indicated mild overall dysphonia (G=0.2) and mild roughness (R=0.2), without evident breathiness, asthenia, or strain. These findings were consistent with mild abnormality in vocal-fold vibration and supported the assigned laryngeal category. \\
& \begin{tabular}[t]{@{}l@{}} \textit{Final}\\ \textit{category}
\end{tabular} &
Benign vocal fold lesions. \\
\midrule
\textbf{Free-text rationale}
& \textit{Instruction} &
Listen carefully to the recording and analyze voice quality, breath use, phonatory stability, pitch, timbre, and abnormal vocal effort. Prioritize directly audible phenotypic features and then integrate the available supporting information to infer the most likely laryngeal category. \\
& \begin{tabular}[t]{@{}l@{}} \textit{Rationale}\\ \textit{target}
\end{tabular} &
The voice showed mild roughness and phonatory instability. Elevated shimmer and noise-to-harmonics ratio were consistent with irregular vocal-fold vibration and an increased noise component in the voice signal. Prolonged, high-intensity occupational voice use provided additional clinical context. Together, the perceptual and acoustic findings were more consistent with a benign vocal-fold lesion than with a primarily functional voice disorder. \\
& \begin{tabular}[t]{@{}l@{}} \textit{Final}\\ \textit{category}
\end{tabular} &
Benign vocal fold lesions. \\
\midrule
\begin{tabular}[t]{@{}l@{}}
\textbf{Empty}\\ \textbf{rationale}
\end{tabular}
& \textit{Instruction} &
Listen to the recording, assess the voice abnormality, and determine the most likely laryngeal category. Provide only the final category without an explanatory rationale. \\
& \begin{tabular}[t]{@{}l@{}} \textit{Rationale}\\ \textit{target}
\end{tabular}
& None. \\ 
& \begin{tabular}[t]{@{}l@{}} \textit{Final}\\ \textit{category}
\end{tabular}
& Benign vocal fold lesions.\\
\bottomrule
\end{tabular}
\footnotetext{
Examples are representative, abridged English translations of Chinese-language training examples from the same underlying case. All three configurations used the same model-input fields and final category target but differed in the formulation of assistant-side rationale supervision. Reference perceptual annotations used to construct rationale targets were not provided as user-side model inputs.
}
\end{table}

\begin{sidewaystable}[p]
\centering
\caption{Configuration of comparator models for seven-class Mandarin provisional laryngeal categorization.}
\label{tab:supp_baseline_configurations}
\small
\setlength{\tabcolsep}{4pt}
\renewcommand{\arraystretch}{1.18}
\begin{tabular}{@{}p{2.5cm}p{4.2cm}p{6.0cm}p{3.8cm}@{}}
\toprule
Model
& Input
& Model configuration
& Training and selection \\
\midrule
SVM
&
1034 handcrafted acoustic descriptors + 9 structured covariates
&
Median imputation and standardization; RBF support-vector classifier
($C=1.0$, $\gamma=\mathrm{scale}$, balanced class weights)
&
Configuration fixed before held-out evaluation
\\
\addlinespace
GMM-i-vector
&
62-dimensional frame-level acoustic features + 9 structured covariates
&
16-component UBM; 50-dimensional length-normalized i-vector;
structured covariates appended before a standardized RBF SVM
($C=10.0$, $\gamma=\mathrm{scale}$, balanced class weights)
&
Configuration fixed before held-out evaluation
\\
\addlinespace
GMM-HMM
&
62-dimensional frame-level acoustic features
&
Class-conditional left-to-right GMM-HMM with 3 states and
2 Gaussian mixtures per state; classification by maximum
length-normalized sequence log-likelihood
&
Configuration fixed before held-out evaluation
\\
\addlinespace
Wav2Vec2-Large
&
Audio only
&
\texttt{facebook/wav2vec2-large}; final-layer encoder states with
attention-mask-aware mean pooling; 7-class linear head; LoRA on query,
key, value, and output attention projections ($r=8$, $\alpha=32$,
dropout 0.05)
&
AdamW; learning rate $1\times10^{-4}$;
batch size 8; maximum 5 epochs
\\
\addlinespace
Whisper-large-v3
&
Audio only
&
\texttt{openai/whisper-large-v3} encoder only; final encoder states
with length-aware mean pooling; 7-class linear head; LoRA on query,
key, value, and output attention projections
($r=8$, $\alpha=32$, dropout 0.05)
&
AdamW; learning rate $1\times10^{-4}$;
batch size 8; maximum 5 epochs
\\
\bottomrule
\end{tabular}
\footnotetext{
The nine encoded structured covariates were age, body-mass index, fundamental
frequency, jitter, shimmer, noise-to-harmonics ratio, maximum phonation time,
and female and male sex indicators. The 62-dimensional frame-level
representation used for the GMM-i-vector and GMM-HMM systems comprised
20 mel-frequency cepstral coefficients, their first- and second-order
derivatives, root-mean-square energy, and zero-crossing rate.
For Wav2Vec2-Large and Whisper-large-v3, the pretrained encoder weights were
otherwise fixed and only the LoRA adapters and classification head were
trainable. Neural baselines used cross-entropy loss, early stopping with a
patience of 2 epochs, and checkpoint selection according to validation-set
weighted F1. UBM denotes universal background model; GMM, Gaussian mixture
model; HMM, hidden Markov model; and LoRA, low-rank adaptation.
}
\end{sidewaystable}

\begin{table}[htbp]
\centering
\caption{Sensitivity of patient-level aggregation to audio-view selection for the model on the 195-patient held-out laryngeal test set.}
\label{tab:supp_aggregation_sensitivity}
\begin{tabular}{@{}lccrrcc@{}}
\toprule
\multirow{2}{*}{Aggregation scheme}
& \multirow{2}{*}{Views, n}
& \multirow{2}{*}{Tied, n}
& \multirow{2}{*}{Accuracy}
& \multirow{2}{*}{Weighted F1}
& \multicolumn{2}{c}{High-risk recognition} \\
\cmidrule(lr){6-7}
& & & & & Sensitivity & Specificity \\
\midrule
Original utterances only
& 586 & 16 & 0.6564 & 0.6543 & 0.6286 & \textbf{0.8625} \\
Single merged view
& 195 & 0 & 0.6718 & 0.6715 & 0.6000 & 0.8562 \\
All views (primary)
& 1795 & 5 & \textbf{0.7179} & \textbf{0.7208} & \textbf{0.7429} & 0.8500 \\
\bottomrule
\end{tabular}
\footnotetext{
All aggregation schemes were evaluated on the same 195 held-out patients using the model.
The single merged view concatenated all available original utterances from the same patient.
The primary analysis used all original and within-patient concatenated audio views.
High-risk sensitivity and specificity treat precancerous or malignant lesions as the positive class under the provisional category alone, not the malignancy alert.
}
\end{table}

\begin{table}[htbp]
\centering
\caption{Sensitivity of patient-level performance to alternative resolutions of tied votes under the primary all-view aggregation scheme.}
\label{tab:supp_tie_sensitivity}
\begin{tabular}{@{}lccc@{}}
\toprule
Metric & Primary analysis & Minimum possible & Maximum possible \\
\midrule
Accuracy & 0.7179 & 0.7026 & 0.7282 \\
Weighted F1 & 0.7208 & 0.7055 & 0.7307 \\
High-risk sensitivity & 0.7429 & 0.7143 & 0.7714 \\
High-risk specificity & 0.8500 & 0.8375 & 0.8500 \\
\bottomrule
\end{tabular}
\footnotetext{
Five held-out patients had tied highest vote counts under the primary all-view aggregation scheme.
The primary analysis retained the seed-controlled tie-breaking rule that was fixed before reader-study card generation.
Minimum and maximum values were obtained across all permissible resolutions of the tied patient-level votes.}
\end{table}

\begin{table}[htbp] 
\centering 
\caption{Class-specific patient-level performance in the temporal-validation cohort.} 
\label{tab:temporal_validation_class_performance} 
\begin{tabular}{@{}lrrrr@{}} 
\toprule Reference category & Participants, n (\%) & Precision & Recall & F1 \\ 
\midrule 
Healthy control & 22 (11.3) & 1.0000 & 0.7727 & 0.8718 \\ 
Benign vocal fold lesions & 63 (32.3) & 0.7656 & 0.7778 & 0.7717 \\ 
Functional dysphonia & 21 (10.8) & 0.6087 & 0.6667 & 0.6364 \\ 
Unilateral vocal fold immobility & 33 (16.9) & 0.5385 & 0.6364 & 0.5833 \\ 
Sulcus vocalis & 12 (6.2) & 0.3077 & 0.3333 & 0.3200 \\ 
Precancerous or malignant lesions & 38 (19.5) & 0.7500 & 0.6316 & 0.6857 \\ 
Spasmodic dysphonia & 6 (3.1) & 0.2857 & 0.3333 & 0.3077 \\ 
\midrule 
Weighted average & 195 & 0.6907 & 0.6718 & 0.6777 \\ 
\bottomrule 
\end{tabular} 
\end{table}

\begin{sidewaystable}[p]
\centering
\caption{Patient-level performance in development-stage task-specific configuration analyses for seven-class Mandarin provisional laryngeal categorization.}
\label{tab:supp_configuration_analysis}
\begin{tabular}{@{}p{3.8cm}p{3.3cm}p{3.6cm}ccc@{}}
\toprule
\multicolumn{3}{c}{Configuration} & \multicolumn{3}{c}{Patient-level performance} \\
\cmidrule(lr){1-3} \cmidrule(lr){4-6}
Rationale supervision & Audio-view setting & Held-out input & Precision & Recall & F1 \\
\midrule
Rule-constrained rationale & Original utterances only & Full input & 0.6584 & 0.6256 & 0.6084 \\
Rule-constrained rationale & Single merged view only & Full input & 0.6201 & 0.6000 & 0.5976 \\
Rule-constrained rationale & All views & Full input & 0.6510 & 0.6564 & 0.6467 \\
Rule-constrained rationale & All views & No objective voice measures & 0.6585 & 0.6410 & 0.6292 \\
Rule-constrained rationale & All views & No patient-context fields & 0.6545 & 0.6359 & 0.6200 \\
Empty rationale & All views & Full input & 0.6647 & 0.6000 & 0.5866 \\
Free-text rationale & All views & Full input & \textbf{0.7075} & \textbf{0.6769} & \textbf{0.6801} \\
\bottomrule
\end{tabular}
\footnotetext{
Values are weighted averages across the seven reference categories.
Full input comprised audio, phonation or speech-task information, available pre-laryngoscopic patient context and objective voice measures.
The no-objective-voice-measures configuration omitted fundamental frequency (F0), jitter, shimmer, noise-to-harmonics ratio (NHR) and maximum phonation time (MPT), while retaining audio, task and patient-context information.
The no-patient-context configuration omitted demographic, occupational and voice-use information, while retaining audio, task information and objective voice measures.
Rule-constrained rationale targets denote clinician-reviewed, rule-constrained acoustic-rationale targets derived from perceptual voice phenotypes and supporting clinical variables.
Free-text rationale targets used the same evidence fields but expressed the rationale in less templated natural language.
Patient-level values were computed using the tie-resolved protocol, in which tied patient-level votes were assigned using a fixed random seed.
The full-input free-text-rationale configuration corresponds to the task-specific auditory large language model reported in Supplementary Table~\ref{tab:laryngeal_etiology_baseline_comparison}.
}
\end{sidewaystable}

\begin{table}[htbp]
\centering
\caption{Clinical reader characteristics in the sequential reader study.}
\label{tab:reader_characteristics}
\small
\begin{tabular}{@{}lc@{}}
\toprule
Characteristic & Readers, n (\%) \\
\midrule
\multicolumn{2}{@{}l}{\textit{Clinical role or professional level}} \\
\quad Postgraduate clinical trainee & 6 (18.8) \\
\quad Resident or junior physician & 10 (31.2) \\
\quad Attending or intermediate physician & 10 (31.2) \\
\quad Chief physician & 6 (18.8) \\
\midrule
\multicolumn{2}{@{}l}{\textit{Experience group}} \\
\quad Less experienced & 16 (50.0) \\
\quad More experienced & 16 (50.0) \\
\midrule
\multicolumn{2}{@{}l}{\textit{Practice region}} \\
\quad Beijing & 18 (56.2) \\
\quad Outside Beijing & 14 (43.8) \\
\bottomrule
\end{tabular}
\footnotetext{
The postgraduate clinical trainees included 2 doctoral and 4 master's trainees.
Less-experienced readers comprised postgraduate clinical trainees and resident or junior physicians; more-experienced readers comprised attending or intermediate physicians and chief physicians.}
\end{table}

\begin{sidewaystable}[p]
\centering
\caption{Additional recording and voice-assessment characteristics of the development and temporal-validation cohorts.}
\label{tab:laryngeal_cohort_clinical_characteristics}
\begin{tabular}{@{}lccccc@{}}
\toprule
 & \multicolumn{4}{c}{Development cohort} & Temporal \\
\cmidrule(lr){2-5}
Characteristic & Training & Validation & Held-out test & Total & validation \\
\midrule
\multicolumn{6}{@{}l}{\textit{Recording availability}} \\
\quad 1 utterance, n (\%) & 152 (9.8) & 20 (10.3) & 19 (9.7) & 191 (9.8) & 22 (11.3) \\
\quad 2 utterances, n (\%) & 23 (1.5) & 1 (0.5) & 4 (2.1) & 28 (1.4) & 1 (0.5) \\
\quad 3 utterances, n (\%) & 1162 (74.7) & 141 (72.3) & 146 (74.9) & 1449 (74.5) & 137 (70.3) \\
\quad $\geq$4 utterances, n (\%) & 219 (14.1) & 33 (16.9) & 26 (13.3) & 278 (14.3) & 35 (17.9) \\
\midrule
\multicolumn{6}{@{}l}{\textit{Patient-reported voice handicap}} \\
\quad VHI-30 total score, median (IQR) & 44.0 (23.0--67.0) & 45.5 (23.0--67.0) & 41.0 (21.8--67.0) & 44.0 (23.0--67.0) & 45.0 (26.0--67.2) \\
\midrule
\multicolumn{6}{@{}l}{\textit{Perceptual and objective voice measures}} \\
\quad GRBAS Grade, median (IQR) & 1.5 (0.8--1.8) & 1.5 (0.8--1.8) & 1.5 (0.8--1.8) & 1.5 (0.8--1.8) & 1.2 (1.0--1.8) \\
\quad F0, Hz, median (IQR) & 196.6 (139.2--231.8) & 192.2 (135.7--235.0) & 203.2 (128.1--234.6) & 197.2 (138.0--232.4) & 200.3 (135.5--236.9) \\
\quad Jitter, \%, median (IQR) & 1.89 (0.93--3.24) & 1.55 (0.83--3.06) & 2.08 (0.97--3.40) & 1.87 (0.92--3.23) & 1.69 (0.95--3.04) \\
\quad Shimmer, \%, median (IQR) & 6.80 (5.00--9.16) & 6.48 (4.62--8.94) & 7.39 (5.09--9.57) & 6.81 (4.96--9.21) & 7.08 (5.23--9.68) \\
\quad NHR, median (IQR) & 0.17 (0.14--0.21) & 0.17 (0.14--0.22) & 0.18 (0.15--0.23) & 0.17 (0.14--0.21) & 0.18 (0.15--0.22) \\
\quad MPT, s, median (IQR) & 10.0 (7.0--15.0) & 11.0 (7.0--15.0) & 10.0 (7.0--15.0) & 10.0 (7.0--15.0) & 9.0 (6.0--12.0) \\
\bottomrule
\end{tabular}
\footnotetext{
Continuous variables are summarized at the patient level as median (interquartile range).
For patients with multiple recordings, GRBAS ratings and objective voice measures were summarized at the patient level before cohort-level aggregation.
VHI-30 denotes 30-item Voice Handicap Index; F0, fundamental frequency; NHR, noise-to-harmonics ratio; and MPT, maximum phonation time.
VHI-30 scores were available for 1,305 training, 156 validation, 164 held-out test, and 164 temporal-validation participants.
Conventional GRBAS Grade ratings were available for 1,521 training, 193 validation, 189 held-out test, and 189 temporal-validation participants; 35, 2, 6, and 6 participants, respectively, had whisper phonation instead of a conventional Grade rating.
F0, jitter, and shimmer were available for 1,497 training, 189 validation, 185 held-out test, and 186 temporal-validation participants; NHR was available for 1,343, 170, 165, and 164 participants, respectively.
MPT was available for all 1,556 training, 195 validation, 195 held-out test, and 195 temporal-validation participants.
}
\end{sidewaystable}

\begin{table}[p]
\centering
\caption{Distribution of development-cohort participants according to clinical assessment year and data split.}
\label{tab:laryngeal_cohort_year_distribution}
\small
\setlength{\tabcolsep}{6pt}
\begin{tabular}{@{}lcccc@{}}
\toprule
Clinical assessment year & Training & Validation & Held-out test & Total \\
\midrule
2004 & 3 (0.2) & 0 (0.0) & 0 (0.0) & 3 (0.2) \\
2006 & 1 (0.1) & 0 (0.0) & 0 (0.0) & 1 (0.1) \\
2012 & 0 (0.0) & 1 (0.5) & 1 (0.5) & 2 (0.1) \\
2013 & 2 (0.1) & 1 (0.5) & 0 (0.0) & 3 (0.2) \\
2014 & 4 (0.3) & 0 (0.0) & 0 (0.0) & 4 (0.2) \\
2015 & 1 (0.1) & 0 (0.0) & 0 (0.0) & 1 (0.1) \\
2018 & 3 (0.2) & 0 (0.0) & 1 (0.5) & 4 (0.2) \\
2019 & 33 (2.1) & 7 (3.6) & 4 (2.1) & 44 (2.3) \\
2020 & 62 (4.0) & 9 (4.6) & 12 (6.2) & 83 (4.3) \\
2021 & 177 (11.4) & 21 (10.8) & 17 (8.7) & 215 (11.0) \\
2022 & 115 (7.4) & 21 (10.8) & 14 (7.2) & 150 (7.7) \\
2023 & 136 (8.7) & 15 (7.7) & 16 (8.2) & 167 (8.6) \\
2024 & 597 (38.4) & 69 (35.4) & 74 (37.9) & 740 (38.0) \\
2025 & 397 (25.5) & 48 (24.6) & 55 (28.2) & 500 (25.7) \\
2026 & 25 (1.6) & 3 (1.5) & 1 (0.5) & 29 (1.5) \\
\midrule
2019 or later & 1542 (99.1) & 193 (99.0) & 193 (99.0) & 1928 (99.1) \\
2024 or later & 1019 (65.5) & 120 (61.5) & 130 (66.7) & 1269 (65.2) \\
\bottomrule
\end{tabular}
\footnotetext{
Values are numbers of participants (percentages within each data split).
Calendar years with no included participants are omitted.
The 2026 row includes clinical assessments through March 2026.
}
\end{table}

\begin{table}[htbp]
\centering
\caption{Sensitivity analyses of the primary reader-study mixed-effects model.}
\label{tab:reader_primary_robustness}
\small
\setlength{\tabcolsep}{5pt}
\begin{tabular}{@{}p{5.0cm}p{6.8cm}c@{}}
\toprule
Analysis & Random-effects specification & \shortstack{Study-round odds ratio\\(95\% CI)} \\
\midrule
Primary model
& Random intercepts for reader and participant
& 1.49 (1.18--1.87) \\
Reader--participant dyad sensitivity model
& Random intercepts for reader, participant, and reader--participant dyad
& 1.92 (1.49--2.49) \\
Reader-slope and dyad sensitivity model
& Reader-specific random intercept and study-round slope, with participant and reader--participant dyad random intercepts
& 1.92 (1.48--2.49) \\
\bottomrule
\end{tabular}
\footnotetext{
The binary outcome was exact concordance of the reader-selected category with the adjudicated seven-class reference category. Study round was the fixed effect of interest. The alternative random-effects specifications were sensitivity analyses of the primary model. All models reported in the table converged and were not singular. Odds ratios from logistic mixed-effects models are conditional on the specified random-effects structure and are therefore not expected to be numerically identical across model specifications.
}
\end{table}

\begin{sidewaystable}[p]
\centering
\caption{Exploratory reader-study performance according to whether the model was correct.}
\label{tab:reader_model_correctness}
\small
\setlength{\tabcolsep}{5pt}
\begin{tabular}{@{}lccccc@{}}
\toprule
Model prediction
& \shortstack{Participants,\\n}
& \shortstack{Paired reader--case\\assessments, n}
& \shortstack{Unaided\\round, \%}
& \shortstack{AI-assisted\\round, \%}
& \shortstack{Absolute difference,\\percentage points (95\% CI)} \\
\midrule
Correct
& 137 & 536 & 55.2 & 70.0
& $+14.7$ (6.8 to 23.6) \\
Incorrect
& 55 & 264 & 38.6 & 29.9
& $-8.7$ ($-18.6$ to $-1.2$) \\
\bottomrule
\end{tabular}
\footnotetext{
The model was considered correct when its final
patient-level category exactly matched the adjudicated seven-class reference category.
Absolute differences were calculated as the AI-assisted round minus the unaided round.
Confidence intervals were estimated using two-way clustered bootstrap resampling by
reader and participant while retaining paired responses across rounds.
The interaction between study round and whether the model was correct yielded an odds ratio of 3.86 (95\% CI, 2.33--6.39; Holm-adjusted $P<0.001$).
In the strata in which the model was correct and incorrect, the corresponding study-round odds ratios were 2.28 (95\% CI, 1.71--3.04) and 0.56 (95\% CI, 0.36--0.86), respectively.}
\end{sidewaystable}

\begin{table}[p]
\centering
\caption{Paired correctness transitions according to whether the model was correct.}
\label{tab:reader_model_correctness_transitions}
\small
\setlength{\tabcolsep}{4pt}
\begin{tabular}{@{}lcccc@{}}
\toprule
Model prediction
& \shortstack{Incorrect in\\both rounds, n (\%)}
& \shortstack{Incorrect-to-\\correct, n (\%)}
& \shortstack{Correct-to-\\incorrect, n (\%)}
& \shortstack{Correct in\\both rounds, n (\%)} \\
\midrule
Correct
& 146 (27.2) & 94 (17.5) & 15 (2.8) & 281 (52.4) \\
Incorrect
& 157 (59.5) & 5 (1.9) & 28 (10.6) & 74 (28.0) \\
\bottomrule
\end{tabular}
\footnotetext{
The model was considered correct when its final patient-level category exactly matched the adjudicated seven-class reference category.
Transitions were defined according to exact concordance of the reader-selected category with the adjudicated seven-class reference category in the unaided and subsequent AI-assisted rounds. Counts represent paired reader--case assessments. Percentages were calculated within strata defined by whether the model was correct. The strata in which the model was correct and incorrect comprised 536 and 264 paired reader--case assessments, respectively.
}
\end{table}

\begin{sidewaystable}[p]
\centering
\caption{Exploratory interaction and behavioral analyses in the reader study.}
\label{tab:reader_exploratory_interactions}
\small
\setlength{\tabcolsep}{5pt}
\begin{tabular}{@{}p{4.6cm}p{3.5cm}ccc p{4.8cm}@{}}
\toprule
Analysis
& \shortstack[l]{Interaction or association\\OR (95\% CI)}
& \shortstack{Likelihood-ratio\\$\chi^2$ (df)}
& \shortstack{Unadjusted\\$P$ value}
& \shortstack{Holm-adjusted\\$P$ value}
& Sensitivity or robustness analysis \\
\midrule
\multicolumn{6}{@{}l}{\textit{Reader-performance heterogeneity}} \\
\quad Study round $\times$ experience group
& 0.90 (0.57--1.42)
& 0.20 (1) & 0.653 & 1.000 & --- \\
\quad Study round $\times$ professional level
& ---
& 2.60 (3) & 0.458 & 1.000 & --- \\
\quad Study round $\times$ professional-level ordinal trend
& 0.88 (0.69--1.11)
& 1.19 (1) & 0.276 & 1.000 & --- \\
\quad Study round $\times$ practice region
& 0.88 (0.55--1.39)
& 0.30 (1) & 0.586 & 1.000 & --- \\
\quad Study round $\times$ reference category
& ---
& 11.83 (6) & 0.066 & 0.396 & --- \\
\quad Study round $\times$ prediction consistency
& 0.97 (0.86--1.09)
& 0.27 (1) & 0.603 & 1.000 & --- \\
\quad Study round $\times$ model correctness
& 3.86 (2.33--6.39)
& 27.46 (1) & $<0.001$ & $<0.001$ & See Supplementary Table~\ref{tab:reader_model_correctness}. \\
\midrule
\multicolumn{6}{@{}l}{\textit{Reader--model concordance heterogeneity}} \\
\quad Study round $\times$ experience group
& 0.61 (0.38--0.97)
& 4.37 (1) & 0.037 & 0.073
& Random-reader-slope LRT, $P=0.263$; reader-level permutation test, $P=0.280$. \\
\quad Study round $\times$ professional level
& ---
& 21.17 (3) & $<0.001$ & $<0.001$
& Random-reader-slope LRT, $P=0.073$; reader-level permutation test, $P=0.095$. \\
\quad Study round $\times$ practice region
& 0.63 (0.39--1.00)
& 3.81 (1) & 0.051 & 0.073 & --- \\
\midrule
\multicolumn{6}{@{}l}{\textit{Response matching after initial reader--model disagreement}} \\
\quad Experience group
& 0.19 (0.03--1.27)
& 2.92 (1) & 0.087 & 0.175 & --- \\
\quad Professional level
& ---
& 9.84 (3) & 0.020 & 0.080 & --- \\
\quad Practice region
& 0.22 (0.03--1.48)
& 2.41 (1) & 0.121 & 0.175 & --- \\
\quad Model correctness
& 1.82 (1.01--3.27)
& 3.88 (1) & 0.049 & 0.147 & --- \\
\bottomrule
\end{tabular}
\footnotetext{
For reader-performance and reader--model-concordance analyses, odds ratios correspond to interaction terms when the subgroup variable was binary or ordinal. A single interaction odds ratio is not reported for categorical variables with more than two levels; heterogeneity for these variables was assessed using the global likelihood-ratio test.
Model correctness indicates whether the model's final patient-level category exactly matched the adjudicated seven-class reference category.
Prediction consistency was analyzed per 0.1 increase in the proportion of valid audio-view predictions assigned to the final patient-level category.
For response-matching analyses, the analysis set was restricted to reader--case assessments for which the unaided reader response differed from the model prediction.
For multiplicity-aware context, Holm adjustment was applied separately within the three exploratory analysis families shown in the table. These adjustments were exploratory and were not part of the primary hypothesis-testing framework.
Reader-level sensitivity analyses treated each of the 32 readers as the independent
observational unit.}
\end{sidewaystable}

\begin{table}[htbp]
\centering
\caption{Patient-level performance for seven-class laryngeal-disease categorization in the held-out test cohort.}
\label{tab:laryngeal_etiology_baseline_comparison}
\begin{tabular}{@{}lccc@{}}
\toprule
Model & Weighted precision & Weighted recall & Weighted F1 \\
\midrule
SVM & 0.5832 & 0.5231 & 0.5323 \\
GMM-i-vector & 0.5139 & 0.4821 & 0.4863 \\
GMM-HMM & 0.4147 & 0.3846 & 0.3495 \\
Wav2Vec2-Large & 0.6648 & 0.5692 & 0.5557 \\
Whisper-large-v3 & 0.7072 & 0.6256 & 0.6180 \\
Task-specific auditory language model & 0.7075 & 0.6769 & 0.6801 \\
Unified model & \textbf{0.7402} & \textbf{0.7179} & \textbf{0.7208} \\
\bottomrule
\end{tabular}
\footnotetext{
Values are weighted averages across the seven reference categories.
The same held-out test cohort and seven-class label set were used for all systems.
Patient-level comparator results are shown graphically in Supplementary Fig.~\ref{fig:supp_laryngeal_standalone}B.
}
\end{table}

\begin{table}[htbp]
\centering
\caption{GRBAS predictive error (MAE) against the senior-reviewed reference ratings.}
\label{tab:grbas_mae}
\begin{tabular}{@{}lcccccc@{}}
\toprule
Model & G & R & B & A & S & Overall \\
\midrule
\multicolumn{7}{@{}l}{\textit{Utterance Level}} \\
\quad SVM & 0.4738 & 0.4698 & 0.3641 & 0.1332 & 0.3308 & 0.3543 \\
\quad GMM-i-vector & 0.4947 & 0.4908 & 0.3762 & 0.1278 & 0.3435 & 0.3666 \\
\quad Wav2Vec2-Large & 0.6380 & 0.7074 & 0.4042 & 0.3451 & 0.3458 & 0.4881 \\
\quad Whisper-large-v3 & \textbf{0.4599} & \textbf{0.4637} & \textbf{0.3521} & 0.2336 & 0.3380 & 0.3695 \\
\quad Task-specific auditory language model & 0.4727 & 0.6106 & 0.4740 & \textbf{0.0692} & 0.1946 & 0.3642 \\
\quad Unified model & 0.4822 & 0.5188 & 0.3981 & 0.0849 & \textbf{0.1772} & \textbf{0.3322} \\
\midrule
\multicolumn{7}{@{}l}{\textit{Patient Level}} \\
\quad SVM & 0.4143 & 0.4345 & 0.3355 & 0.1083 & 0.1701 & 0.2925 \\
\quad GMM-i-vector & 0.4276 & 0.4414 & 0.3193 & 0.1118 & 0.1675 & 0.2935 \\
\quad Wav2Vec2-Large & 0.5519 & 0.5352 & 0.3744 & 0.3644 & 0.2128 & 0.4078 \\
\quad Whisper-large-v3 & 0.4100 & \textbf{0.3958} & \textbf{0.3184} & 0.2291 & 0.2545 & 0.3216 \\
\quad Task-specific auditory language model & 0.3581 & 0.4861 & 0.4114 & \textbf{0.0786} & 0.0861 & 0.2841 \\
\quad Unified model & \textbf{0.3476} & 0.4111 & 0.3437 & 0.0889 & \textbf{0.0685} & \textbf{0.2519} \\
\bottomrule
\end{tabular}
\footnotetext{
G denotes Grade, R Roughness, B Breathiness, A Asthenia, and S Strain.
Lower values indicate smaller prediction error.
}
\end{table}

\begin{table}[htbp]
\centering
\caption{GRBAS intraclass correlation coefficients (ICC) with the senior-reviewed reference ratings.}
\label{tab:grbas_icc}
\begin{tabular}{@{}lcccccc@{}}
\toprule
Model & G & R & B & A & S & Overall \\
\midrule
\multicolumn{7}{@{}l}{\textit{Utterance Level}} \\
\quad SVM & 0.4756 & 0.4215 & 0.6186 & 0.3714 & 0.4647 & 0.4704 \\
\quad GMM-i-vector & 0.4596 & 0.3781 & 0.5951 & 0.4288 & 0.3257 & 0.4375 \\
\quad Wav2Vec2-Large & 0.4619 & 0.3964 & 0.6086 & 0.1486 & 0.3884 & 0.4008 \\
\quad Whisper-large-v3 & \textbf{0.6567} & \textbf{0.5447} & \textbf{0.6928} & 0.3015 & 0.6509 & 0.5693 \\
\quad Task-specific auditory language model & 0.6425 & 0.4698 & 0.5395 & \textbf{0.4955} & 0.7112 & 0.5717 \\
\quad Unified model & 0.6223 & 0.4835 & 0.6208 & 0.4647 & \textbf{0.7543} & \textbf{0.5891} \\
\midrule
\multicolumn{7}{@{}l}{\textit{Patient Level}} \\
\quad SVM & 0.6470 & 0.5437 & 0.6933 & 0.4621 & 0.5419 & 0.5776 \\
\quad GMM-i-vector & 0.6296 & 0.5386 & 0.7264 & 0.5119 & 0.4592 & 0.5731 \\
\quad Wav2Vec2-Large & 0.5445 & 0.5457 & 0.6945 & 0.1719 & 0.3506 & 0.4614 \\
\quad Whisper-large-v3 & 0.7336 & 0.6669 & \textbf{0.7826} & 0.3891 & 0.5895 & 0.6323 \\
\quad Task-specific auditory language model & 0.7801 & 0.6152 & 0.6504 & 0.5791 & 0.7235 & 0.6697 \\
\quad Unified model & \textbf{0.7991} & \textbf{0.6720} & 0.7382 & \textbf{0.6327} & \textbf{0.8039} & \textbf{0.7292} \\
\bottomrule
\end{tabular}
\footnotetext{
G denotes Grade, R Roughness, B Breathiness, A Asthenia, and S Strain.
Higher values indicate stronger agreement with the reference ratings.
}
\end{table}

\begin{table}[htbp]
\centering
\caption{GRBAS Pearson correlation coefficients ($r$) with the senior-reviewed reference ratings.}
\label{tab:grbas_pearson}
\begin{tabular}{@{}lcccccc@{}}
\toprule
Model & G & R & B & A & S & Overall \\
\midrule
\multicolumn{7}{@{}l}{\textit{Utterance Level}} \\
\quad SVM & 0.5486 & 0.4784 & 0.6397 & 0.3958 & 0.6665 & 0.5458 \\
\quad GMM-i-vector & 0.5074 & 0.4232 & 0.6151 & 0.4417 & 0.4606 & 0.4896 \\
\quad Wav2Vec2-Large & 0.5793 & 0.4562 & 0.6133 & 0.2490 & 0.6867 & 0.5169 \\
\quad Whisper-large-v3 & \textbf{0.6590} & \textbf{0.5460} & \textbf{0.7028} & 0.4317 & 0.7489 & \textbf{0.6177} \\
\quad Task-specific auditory language model & 0.6464 & 0.4836 & 0.5849 & \textbf{0.5331} & 0.7367 & 0.5969 \\
\quad Unified model & 0.6235 & 0.4843 & 0.6519 & \textbf{0.5223} & \textbf{0.7733} & 0.6111 \\
\midrule
\multicolumn{7}{@{}l}{\textit{Patient Level}} \\
\quad SVM & 0.7382 & 0.6493 & 0.7367 & 0.5312 & 0.6727 & 0.6656 \\
\quad GMM-i-vector & 0.6968 & 0.6108 & 0.7644 & 0.5539 & 0.5234 & 0.6299 \\
\quad Wav2Vec2-Large & 0.6848 & 0.5941 & 0.7098 & 0.3507 & \textbf{0.8488} & 0.6376 \\
\quad Whisper-large-v3 & 0.7469 & 0.6842 & \textbf{0.7924} & 0.5414 & 0.7431 & 0.7016 \\
\quad Task-specific auditory language model & 0.7813 & 0.6197 & 0.6812 & 0.6019 & 0.7248 & 0.6818 \\
\quad Unified model & \textbf{0.7995} & \textbf{0.6867} & 0.7574 & \textbf{0.6669} & 0.8047 & \textbf{0.7430} \\
\bottomrule
\end{tabular}
\footnotetext{
G denotes Grade, R Roughness, B Breathiness, A Asthenia, and S Strain.
Higher positive values indicate stronger linear association with the reference ratings.
}
\end{table}

\begin{table}[htbp]
\centering
\caption{GRBAS Spearman rank correlation coefficients ($\rho$) with the senior-reviewed reference ratings.}
\label{tab:grbas_spearman}
\begin{tabular}{@{}lcccccc@{}}
\toprule
Model & G & R & B & A & S & Overall \\
\midrule
\multicolumn{7}{@{}l}{\textit{Utterance Level}} \\
\quad SVM & 0.5290 & 0.4240 & 0.5630 & 0.3066 & 0.3699 & 0.4385 \\
\quad GMM-i-vector & 0.5043 & 0.3979 & 0.4915 & 0.3204 & 0.3822 & 0.4193 \\
\quad Wav2Vec2-Large & 0.5929 & 0.4245 & 0.5326 & 0.2428 & 0.4808 & 0.4547 \\
\quad Whisper-large-v3 & \textbf{0.6483} & \textbf{0.4943} & 0.6150 & 0.3162 & 0.4523 & 0.5052 \\
\quad Task-specific auditory language model & 0.6444 & 0.4532 & 0.5900 & \textbf{0.5533} & 0.6881 & \textbf{0.5858} \\
\quad Unified model & 0.6118 & 0.4470 & \textbf{0.6361} & 0.5164 & \textbf{0.7039} & 0.5830 \\
\midrule
\multicolumn{7}{@{}l}{\textit{Patient Level}} \\
\quad SVM & 0.7070 & 0.5913 & 0.6451 & 0.3666 & 0.3185 & 0.5257 \\
\quad GMM-i-vector & 0.6954 & 0.5775 & 0.6513 & 0.3739 & 0.3155 & 0.5227 \\
\quad Wav2Vec2-Large & 0.7030 & 0.6127 & 0.5325 & 0.3242 & 0.3851 & 0.5115 \\
\quad Whisper-large-v3 & 0.7322 & \textbf{0.6391} & 0.6736 & 0.3658 & 0.3362 & 0.5494 \\
\quad Task-specific auditory language model & 0.7434 & 0.5822 & 0.6779 & \textbf{0.5123} & 0.5528 & 0.6137 \\
\quad Unified model & \textbf{0.7736} & 0.6185 & \textbf{0.7180} & 0.4585 & \textbf{0.6315} & \textbf{0.6400} \\
\bottomrule
\end{tabular}
\footnotetext{
G denotes Grade, R Roughness, B Breathiness, A Asthenia, and S Strain.
Higher positive values indicate stronger rank association with the reference ratings.
}
\end{table}

\begin{table}[htbp]
\centering
\caption{Mean absolute error (MAE) for VHI-30 score estimation in the held-out cohort.}
\label{tab:vhi_mae}
\begin{tabular}{@{}lccccc@{}}
\toprule
Model & VHI-E & VHI-F & VHI-P & VHI-Total & VHI-Sum \\
\midrule
\multicolumn{6}{@{}l}{\textit{Utterance Level}} \\
\quad SVM & 9.0665 & 7.4334 & 7.6569 & 21.6709 & 21.7658 \\
\quad GMM-i-vector & 9.1598 & \textbf{7.2740} & 7.6518 & 21.7002 & 21.5331 \\
\quad Wav2Vec2-Large & 9.2842 & 7.2964 & 8.3719 & 22.2492 & \textbf{20.6862} \\
\quad Whisper-large-v3 & \textbf{8.8058} & 7.5676 & \textbf{7.2103} & \textbf{21.1139} & 20.7624 \\
\quad Task-specific auditory language model & 9.5316 & 8.2852 & 7.7017 & 27.8007 & 21.7643 \\
\quad Unified model & 9.3580 & 7.8813 & 8.1635 & 23.8019 & 21.7560 \\
\midrule
\multicolumn{6}{@{}l}{\textit{Patient Level}} \\
\quad SVM & 9.0506 & 7.5234 & 7.5167 & 22.1175 & 21.6448 \\
\quad GMM-i-vector & 8.6786 & 7.2505 & 7.4140 & 20.9690 & 20.9342 \\
\quad Wav2Vec2-Large & 9.4884 & 7.5800 & 8.2145 & 22.5415 & 20.5419 \\
\quad Whisper-large-v3 & 8.6687 & 7.7416 & 7.2884 & \textbf{19.9051} & 21.2726 \\
\quad Task-specific auditory language model & 8.7210 & 8.0524 & \textbf{7.1741} & 23.0170 & 20.9000 \\
\quad Unified model & \textbf{8.6490} & \textbf{6.7660} & 7.4252 & 20.5050 & \textbf{20.0512} \\
\bottomrule
\end{tabular}
\footnotetext{
Patient-level VHI-30 analyses included the 164 held-out participants with
available reference VHI-30 scores.
E denotes Emotional, F Functional, and P Physical. VHI-Total denotes the directly estimated VHI-30 total score. VHI-Sum denotes the sum of the estimated Emotional, Functional, and Physical subscale scores evaluated against the reference VHI-30 total score.
Lower values indicate smaller prediction error.
}
\end{table}

\begin{table}[htbp]
\centering
\caption{Intraclass correlation coefficients (ICC) between model-estimated and reference VHI-30 scores in the held-out cohort.}
\label{tab:vhi_icc}
\begin{tabular}{@{}lccccc@{}}
\toprule
Model & VHI-E & VHI-F & VHI-P & VHI-Total & VHI-Sum \\
\midrule
\multicolumn{6}{@{}l}{\textit{Utterance Level}} \\
\quad SVM & 0.2227 & 0.4006 & 0.3281 & 0.3594 & 0.3609 \\
\quad GMM-i-vector & 0.2672 & 0.4511 & 0.3696 & 0.4293 & 0.4230 \\
\quad Wav2Vec2-Large & 0.1845 & 0.5087 & 0.3393 & 0.3340 & 0.4372 \\
\quad Whisper-large-v3 & 0.2511 & 0.5033 & \textbf{0.4175} & \textbf{0.4985} & 0.4626 \\
\quad Task-specific auditory language model & \textbf{0.2951} & 0.4242 & 0.3485 & 0.4253 & 0.4430 \\
\quad Unified model & 0.2870 & \textbf{0.5321} & 0.4057 & 0.4765 & \textbf{0.4909} \\
\midrule
\multicolumn{6}{@{}l}{\textit{Patient Level}} \\
\quad SVM & 0.2156 & 0.3958 & 0.3223 & 0.3229 & 0.3559 \\
\quad GMM-i-vector & 0.3246 & 0.4793 & 0.3921 & 0.4680 & 0.4520 \\
\quad Wav2Vec2-Large & 0.1188 & 0.4100 & 0.3026 & 0.2587 & 0.3579 \\
\quad Whisper-large-v3 & 0.2792 & 0.4781 & 0.4011 & 0.5157 & 0.4428 \\
\quad Task-specific auditory language model & 0.3451 & 0.4423 & 0.3968 & 0.5129 & 0.4650 \\
\quad Unified model & \textbf{0.3684} & \textbf{0.5755} & \textbf{0.4379} & \textbf{0.5340} & \textbf{0.5279} \\
\bottomrule
\end{tabular}
\footnotetext{
Patient-level VHI-30 analyses included the 164 held-out participants with
available reference VHI-30 scores.
E denotes Emotional, F Functional, and P Physical. VHI-Total denotes the directly estimated VHI-30 total score. VHI-Sum denotes the sum of the estimated Emotional, Functional, and Physical subscale scores evaluated against the reference VHI-30 total score.
Higher values indicate stronger agreement with the reference ratings.
}
\end{table}

\begin{table}[htbp]
\centering
\caption{Pearson correlation coefficients ($r$) between model-estimated and reference VHI-30 scores in the held-out cohort.}
\label{tab:vhi_pearson}
\begin{tabular}{@{}lccccc@{}}
\toprule
Model & VHI-E & VHI-F & VHI-P & VHI-Total & VHI-Sum \\
\midrule
\multicolumn{6}{@{}l}{\textit{Utterance Level}} \\
\quad SVM & 0.3351 & 0.4829 & 0.3940 & 0.4524 & 0.4491 \\
\quad GMM-i-vector & 0.3001 & 0.4852 & 0.4095 & 0.4582 & 0.4659 \\
\quad Wav2Vec2-Large & 0.2781 & 0.5346 & 0.3890 & 0.3923 & 0.4856 \\
\quad Whisper-large-v3 & 0.3244 & \textbf{0.6092} & \textbf{0.4492} & \textbf{0.5115} & \textbf{0.5381} \\
\quad Task-specific auditory language model & \textbf{0.3457} & 0.4655 & 0.3722 & 0.4395 & 0.4866 \\
\quad Unified model & 0.3079 & 0.5361 & 0.4081 & 0.4799 & 0.4993 \\
\midrule
\multicolumn{6}{@{}l}{\textit{Patient Level}} \\
\quad SVM & 0.3821 & 0.5303 & 0.4155 & 0.4712 & 0.4913 \\
\quad GMM-i-vector & 0.3856 & 0.5328 & 0.4551 & 0.5153 & 0.5125 \\
\quad Wav2Vec2-Large & 0.3134 & 0.4922 & 0.3971 & 0.4370 & 0.4818 \\
\quad Whisper-large-v3 & 0.4181 & \textbf{0.6222} & \textbf{0.4619} & \textbf{0.5456} & \textbf{0.5602} \\
\quad Task-specific auditory language model & \textbf{0.4644} & 0.5349 & 0.4577 & 0.5140 & 0.5485 \\
\quad Unified model & 0.4304 & 0.5975 & 0.4525 & 0.5338 & 0.5527 \\
\bottomrule
\end{tabular}
\footnotetext{
Patient-level VHI-30 analyses included the 164 held-out participants with
available reference VHI-30 scores.
E denotes Emotional, F Functional, and P Physical. VHI-Total denotes the directly estimated VHI-30 total score. VHI-Sum denotes the sum of the estimated Emotional, Functional, and Physical subscale scores evaluated against the reference VHI-30 total score.
Higher positive values indicate stronger linear association with the reference ratings.
}
\end{table}

\begin{table}[htbp]
\centering
\caption{Spearman rank correlation coefficients ($\rho$) between model-estimated and reference VHI-30 scores in the held-out cohort.}
\label{tab:vhi_spearman}
\begin{tabular}{@{}lccccc@{}}
\toprule
Model & VHI-E & VHI-F & VHI-P & VHI-Total & VHI-Sum \\
\midrule
\multicolumn{6}{@{}l}{\textit{Utterance Level}} \\
\quad SVM & 0.3130 & 0.4722 & 0.4007 & 0.4461 & 0.4426 \\
\quad GMM-i-vector & 0.3200 & 0.4927 & 0.4081 & 0.4645 & 0.4825 \\
\quad Wav2Vec2-Large & 0.3277 & 0.4977 & 0.4106 & 0.3994 & 0.5057 \\
\quad Whisper-large-v3 & 0.3608 & \textbf{0.6180} & \textbf{0.4487} & \textbf{0.5250} & \textbf{0.5557} \\
\quad Task-specific auditory language model & \textbf{0.3743} & 0.4795 & 0.3766 & 0.4269 & 0.5010 \\
\quad Unified model & 0.3270 & 0.5500 & 0.4140 & 0.4784 & 0.5114 \\
\midrule
\multicolumn{6}{@{}l}{\textit{Patient Level}} \\
\quad SVM & 0.3384 & 0.4932 & 0.4266 & 0.4509 & 0.4668 \\
\quad GMM-i-vector & 0.3969 & 0.5331 & 0.4623 & 0.5112 & 0.5203 \\
\quad Wav2Vec2-Large & 0.3498 & 0.5015 & 0.4390 & 0.4166 & 0.5194 \\
\quad Whisper-large-v3 & 0.4295 & \textbf{0.6201} & \textbf{0.4784} & \textbf{0.5588} & 0.5668 \\
\quad Task-specific auditory language model & \textbf{0.4511} & 0.4850 & 0.4718 & 0.5297 & 0.5503 \\
\quad Unified model & 0.4395 & 0.5972 & 0.4679 & 0.5525 & \textbf{0.5710} \\
\bottomrule
\end{tabular}
\footnotetext{
Patient-level VHI-30 analyses included the 164 held-out participants with
available reference VHI-30 scores.
E denotes Emotional, F Functional, and P Physical. VHI-Total denotes the directly estimated VHI-30 total score. VHI-Sum denotes the sum of the estimated Emotional, Functional, and Physical subscale scores evaluated against the reference VHI-30 total score.
Higher positive values indicate stronger rank association with the reference ratings.
}
\end{table}

\begin{table}[htbp]
\centering
\caption{Performance for categorical and continuous pitch assessment in the held-out cohort.}
\label{tab:pitch_results}
\begin{tabular}{@{}lccccc@{}}
\toprule
Model & MAE & Acc & Pearson & Spearman & ICC \\
\midrule
\multicolumn{6}{@{}l}{\textit{Utterance Level}} \\
\quad SVM & 0.4388 & 0.4312 & 0.5939 & 0.5862 & 0.5570 \\
\quad GMM-i-vector & 0.4980 & 0.5536 & 0.5756 & 0.5315 & 0.5617 \\
\quad Wav2Vec2-Large & 0.6023 & 0.2684 & 0.3476 & 0.3490 & 0.3089 \\
\quad Whisper-large-v3 & 0.4866 & 0.3616 & \textbf{0.6129} & \textbf{0.6310} & \textbf{0.6106} \\
\quad Task-specific auditory language model & 0.4793 & \textbf{0.6412} & 0.4553 & 0.4360 & 0.4337 \\
\quad Unified model & 0.5154 & 0.5806 & 0.5341 & 0.5359 & 0.5283 \\
\midrule
\multicolumn{6}{@{}l}{\textit{Patient Level}} \\
\quad SVM & 0.4314 & 0.4456 & 0.6113 & 0.6170 & 0.5215 \\
\quad GMM-i-vector & 0.4645 & 0.5181 & 0.5758 & 0.5484 & 0.5471 \\
\quad Wav2Vec2-Large & 0.5066 & 0.2487 & 0.5162 & 0.5215 & 0.3436 \\
\quad Whisper-large-v3 & 0.4389 & 0.6218 & \textbf{0.6455} & \textbf{0.6667} & 0.6211 \\
\quad Task-specific auditory language model & 0.4349 & 0.6373 & 0.4826 & 0.5189 & 0.4326 \\
\quad Unified model & \textbf{0.4311} & \textbf{0.6736} & 0.6292 & 0.6334 & \textbf{0.6285} \\
\bottomrule
\end{tabular}
\footnotetext{
Accuracy was calculated across the three pitch-status categories of normal,
raised, and lowered pitch. Mean absolute error (MAE), Pearson correlation,
Spearman rank correlation, and intraclass correlation coefficient (ICC) were
calculated from the continuous signed pitch values, with positive and negative
values indicating raised and lowered pitch, respectively, relative to the
neutral reference. For pretrained speech baselines without explicit
categorical pitch outputs, categories were derived from the signed continuous
pitch predictions using zero as the neutral threshold.
}
\end{table}

\begin{table}[p]
\centering
\caption{Disease-specific reader performance in the unaided and subsequent AI-assisted rounds.}
\label{tab:reader_disease_specific}
\footnotesize
\setlength{\tabcolsep}{3pt}
\begin{tabular}{@{}p{2.6cm}clccc@{}}
\toprule
Reference category
& \shortstack{Paired\\assessments,\\n (\%)}
& Metric
& \shortstack{Unaided\\(\%)}
& \shortstack{AI-assisted\\(\%)}
& \shortstack{Difference, pp\\(95\% CI)} \\
\midrule
\multirow{4}{2.6cm}{Healthy control}
& \multirow{4}{*}{77 (9.6)}
& Recall      & 62.3 & 61.0 & $-1.3$ ($-11.9$ to 9.0) \\
& & Precision   & 32.4 & 45.2 & 12.8 (3.6 to 25.4) \\
& & F1          & 42.7 & 51.9 & 9.3 (0.9 to 19.3) \\
& & Specificity & 86.2 & 92.1 & 5.9 (2.1 to 11.1) \\
\addlinespace
\multirow{4}{2.6cm}{Benign vocal fold lesions}
& \multirow{4}{*}{253 (31.6)}
& Recall      & 54.5 & 56.1 & 1.6 ($-8.9$ to 11.7) \\
& & Precision   & 53.7 & 58.7 & 5.0 ($-1.9$ to 12.6) \\
& & F1          & 54.1 & 57.4 & 3.3 ($-4.9$ to 11.4) \\
& & Specificity & 78.2 & 81.7 & 3.5 ($-1.0$ to 8.1) \\
\addlinespace
\multirow{4}{2.6cm}{Functional dysphonia}
& \multirow{4}{*}{85 (10.6)}
& Recall      & 42.4 & 47.1 & 4.7 ($-12.3$ to 23.9) \\
& & Precision   & 45.0 & 52.6 & 7.6 ($-6.4$ to 24.2) \\
& & F1          & 43.6 & 49.7 & 6.1 ($-8.6$ to 22.6) \\
& & Specificity & 93.8 & 95.0 & 1.1 ($-1.4$ to 3.7) \\
\addlinespace
\multirow{4}{2.6cm}{Unilateral vocal fold immobility}
& \multirow{4}{*}{128 (16.0)}
& Recall      & 39.1 & 46.1 & 7.0 ($-3.8$ to 19.7) \\
& & Precision   & 64.9 & 64.8 & $-0.1$ ($-13.9$ to 13.2) \\
& & F1          & 48.8 & 53.9 & 5.1 ($-5.5$ to 16.2) \\
& & Specificity & 96.0 & 95.2 & $-0.7$ ($-3.2$ to 1.5) \\
\addlinespace
\multirow{4}{2.6cm}{Sulcus vocalis}
& \multirow{4}{*}{93 (11.6)}
& Recall      & 36.6 & 47.3 & 10.8 ($-1.8$ to 26.2) \\
& & Precision   & 63.0 & 71.0 & 8.0 ($-8.6$ to 26.6) \\
& & F1          & 46.3 & 56.8 & 10.5 ($-2.2$ to 25.7) \\
& & Specificity & 97.2 & 97.5 & 0.3 ($-1.6$ to 2.1) \\
\addlinespace
\multirow{4}{2.6cm}{Precancerous or malignant lesions}
& \multirow{4}{*}{128 (16.0)}
& Recall      & 53.9 & 75.0 & 21.1 (4.3 to 42.0) \\
& & Precision   & 52.3 & 54.5 & 2.3 ($-9.5$ to 16.0) \\
& & F1          & 53.1 & 63.2 & 10.1 ($-1.8$ to 25.1) \\
& & Specificity & 90.6 & 88.1 & $-2.5$ ($-7.3$ to 2.0) \\
\addlinespace
\multirow{4}{2.6cm}{Spasmodic dysphonia}
& \multirow{4}{*}{36 (4.5)}
& Recall      & 63.9 & 72.2 & 8.3 ($-25.0$ to 41.7) \\
& & Precision   & 44.2 & 53.1 & 8.8 ($-9.4$ to 31.2) \\
& & F1          & 52.3 & 61.2 & 8.9 ($-13.4$ to 32.5) \\
& & Specificity & 96.2 & 97.0 & 0.8 ($-1.0$ to 3.0) \\
\bottomrule
\end{tabular}
\footnotetext{
Paired reader--case assessments reflect all 800 assessments from 32 readers. Each reader evaluated the same randomly assigned cases in the unaided and AI-assisted rounds.
Unaided and AI-assisted values are percentages. Differences are expressed in percentage points and were calculated as AI-assisted minus unaided performance. Recall corresponds to class-specific sensitivity, and specificity was calculated one-versus-rest. Confidence intervals for between-round differences were estimated using two-way clustered bootstrap resampling at the reader and participant levels while preserving the paired round observations.
}
\end{table}


\begin{table}[htbp]
\centering
\caption{Patient-level performance for four-class dysarthria etiologic categorization in the held-out cohort.}
\label{tab:etiology_prediction}
\begin{tabular}{@{}lccc@{}}
\toprule
Model & Weighted precision & Weighted recall & Weighted F1 \\
\midrule
SVM & 0.7873 & 0.7451 & 0.7539 \\
GMM-i-vector & 0.7156 & 0.5882 & 0.6386 \\
GMM-HMM & 0.6198 & 0.5098 & 0.5558 \\
Wav2Vec2-Large & 0.7484 & 0.7059 & 0.7048 \\
Whisper-large-v3 & \textbf{0.8868} & 0.7647 & 0.8002 \\
Task-specific auditory language model & 0.8486 & 0.8039 & 0.8191 \\
Unified model & 0.8827 & \textbf{0.8627} & \textbf{0.8686} \\
\bottomrule
\end{tabular}
\footnotetext{Values are weighted averages across Parkinson's disease, cerebral palsy, amyotrophic lateral sclerosis and Down syndrome.}
\end{table}

\end{document}